\documentclass[aps,pra,twocolumn,showpacs,10pt]{revtex4-1}

\usepackage{amsmath,amssymb}
\usepackage{graphicx}

\graphicspath{ {.} {./pics/} }

\newcommand{\rme}{\mathrm e}
\newcommand{\rmi}{\mathrm i}
\newcommand{\rmd}{{\mathrm d}}

\renewcommand{\Re}{\mathop{{\rm Re}}}
\renewcommand{\Im}{\mathop{{\rm Im}}}

\begin{document}

\title{Nonclassical light from few emitters in a cavity}

\author{D. Pagel}
\author{A. Alvermann}
\author{H. Fehske}
\affiliation{Institut f\"ur Physik, Ernst-Moritz-Arndt-Universit\"at, 17487 Greifswald, Germany}

\begin{abstract}
We study the characteristics of the light generated by few
emitters in a cavity at strong light-matter coupling.
By means of the Glauber $g^{(2)}$-function
we can identify clearly distinguished parameter regimes with
super-Poissonian and sub-Poissonian photon statistics.
We establish 
a relation between the emission characteristics for one and multiple emitters,
and explain its origin in terms of the photon-dressed emitter states.
Cooperative effects lead to the generation of
nonclassical light already at reduced light-matter coupling if the number of emitters is increased.
Our results are obtained with a full input-output formalism and master equation valid also at strong light-matter coupling.
We compare the behavior obtained with and without counter-rotating light-matter interaction terms in the Hamiltonian, and find that the generation of nonclassical light is robust against such modifications.
Finally, we contrast our findings with the predictions of the quantum optical master equation and find that it fails entirely
at predicting regimes with different photon statistics.
\end{abstract}


\maketitle

\section{Introduction}

Two-level emitters interacting with a cavity photon mode are widely studied in quantum optics with respect to spontaneous emission and superradiance~\cite{BSH71, GH82, MH10, BERB12}, cooperativity and lasing~\cite{AGPDS11, LLdVFVB11, MGSA13},
as well as the emission of nonclassical light~\cite{AT91, TW09, Que12}.
For sufficiently weak light-matter coupling,
when the photon-dressing of emitter states is negligibly small, the emitter-cavity system can be studied with the quantum optical master equation, usually in combination with the rotating-wave approximation~\cite{Car99}.
The quantum optical master equation fails at strong light-matter coupling,
to the extent that it predicts unphysical emission at zero temperature if the number of photons in the ground state is finite~\cite{CBC05, CC06,WDDVB08}.

The correct theoretical description of systems with (ultra-)strong light-matter coupling~\cite{TRK92, MNBBBK05, SFSLR13, GP13} has attracted increasing interest recently~\cite{RLSH12, SRDSHS13, GRSDSS13, RSH13, PFSV13, BGB11}.
Essentially, the quantum optical master equation has to be replaced by a master equation expressed in the photon-dressed emitter eigenstates~\cite{BP02,ALZ06,SB08,BGB11,Sch14,Agar13}. While the master equation remains Markovian, which is justified because of the weak emitter-environment and cavity-environment couplings~\cite{BP02, Car99},
it now requires full diagonalization of the interacting emitter-cavity Hamiltonian.
Such an equation was used in recent studies of photon blockade effects~\cite{RLSH12}, spontaneous conversion of virtual to real photons~\cite{SRDSHS13, GRSDSS13}, and the emission of nonclassical light from a single emitter~\cite{RSH13}.

In this paper we study the emission of few emitters in a cavity, with particular focus on the photon statistics of the emitted light. Our goal is the characterization of temperature and coupling regimes 
where nonclassical light~\cite{Sin83, MW95} is generated.
A major result will be the identification of two clearly distinguished neighboring regimes with pronounced sub-Poissonian and super-Poissonian photon statistics at strong coupling.

Our results are obtained with the full input-output formalism~\cite{CG84, GC85, Gra89, SG96pra} and master equation~\cite{BP02,ALZ06,SB08,BGB11,Sch14,Agar13} without further approximations.
To understand the relevance of the different approximations involved in traditional quantum optics treatments we make two comparisons.
First, we compare the results that are obtained when the counter-rotating light-matter interaction terms are included in the Hamiltonian to those when they are dropped.
Second, we contrast the results obtained with the full master equation with results from the quantum optical master equation.
The latter comparison will clearly show the necessity of using the correct master equation already at weak coupling if the photon statistics is of interest.
This issue has been studied conclusively for a single emitter in Ref.~\cite{RSH13},
which also contains Glauber function plots for few emitters in the supplemental material but omits the further analysis of the situation that we give here.

The paper is organized as follows.
In Sec.\ \ref{sec:model} we introduce the physical situation under study
together with the master equation used for its analysis.
In Sec.\ \ref{sec:spectra} we discuss the emission spectra in relation to the energy spectra of the emitter-cavity Hamiltonian,
while the statistics of the emitted photons is studied in Sec.\ \ref{sec:statistics}.
We conclude in Sec.\ \ref{sec:concl}.
The appendices collect further information on the theoretical approach.
App.\ \ref{app:inout} gives details of the input-output formalism.
In App.\ \ref{app:ME} we derive the master equation,
and give a few analytical results for the photon statistics in App.\ \ref{app:analyt}.

\section{\label{sec:model}The physical situation}
The interaction of $N$ two-level emitters with a single cavity photon mode is described by the Dicke model~\cite{Dic54}
\begin{multline}\label{HS}
 H = \omega_c a^\dagger a  + \omega_x \sum_{j=1}^N \sigma_+^{(j)} \sigma_-^{(j)}
  + g \sum_{j=1}^N (a^\dagger\sigma_-^{(j)} + a \sigma_+^{(j)}) \\ +
  g'  \sum_{j=1}^N (a \sigma_-^{(j)} + a^\dagger \sigma_+^{(j)})  \,, \qquad
\end{multline}
where the operator $a^{(\dagger)}$ annihilates (creates) a cavity photon with frequency  $\omega_c$ and $\sigma_-^{(j)}$ ($\sigma_+^{(j)}$) is the corresponding lowering (raising) operator for the $j$th emitter with transition energy $\omega_x$.
Throughout this work, we consider the resonant case $\omega_0 = \omega_c = \omega_x$.
We allow for different emitter-photon coupling strengths
for the co-rotating ($g$) and counter-rotating ($g'$) interaction terms.
Changing $g'$ relative to $g$ interpolates between the Tavis-Cummings (TC) limit ($g'=0$) without and the Dicke limit ($g'=g$) with counter-rotating terms. 
Both situations can be realized experimentally~\cite{DEPC07, BAGPB14}.
The rotating-wave approximation consists in replacing the Dicke by the TC limit.

Dissipation arises from the coupling of the emitters and the cavity to the environment.
For a bosonic environment the coupling terms are of the form 
\begin{equation}\label{HI}
 H_I = -\rmi S \sum_\nu \lambda_\nu (b_\nu - b_\nu^\dagger) \,,
\end{equation}
where $S$ is a (Hermitian) emitter or cavity operator
and the $b_\nu^{(\dagger)}$ are bosonic operators for the environment photons (at frequencies $\omega_\nu$ with coupling constants $\lambda_\nu$).
As the operator $S$ we choose the field operator $X = -\rmi X_0 (a - a^\dagger)$ for the coupling of the cavity and the transition operator $\sigma_y^{(j)} = \rmi (\sigma_+^{(j)} - \sigma_-^{(j)})$ for the coupling of the $j$th emitter to the environment.

At sufficiently weak coupling to the environment,
the emitter-cavity system density matrix $\rho$ obeys a Markovian master equation~\cite{BP02,ALZ06,SB08,BGB11,Sch14,Agar13}
\begin{eqnarray}\label{ME}
 \frac{\rmd}{\rmd t} \rho(t) &=& -\rmi [H, \rho(t)] - \rmi \frac{1}{2} \sum_\omega \xi(\omega) \big[ S_\omega^\dagger S_\omega, \rho(t) \big] \nonumber\\
 && + \frac{1}{2} \sum_\omega \chi(\omega) \Big( \big[ S_\omega \rho(t), S_\omega^\dagger \big] + \big[ S_\omega, \rho(t) S_\omega^\dagger \big] \Big) \,, \nonumber\\
\end{eqnarray}
where
\begin{equation}\label{S}
 S_\omega = \sum_{m,n} | m \rangle \langle m | S | n \rangle \langle n | \delta_{E_n - E_m, \omega}
\end{equation}
is the projection of $S$ onto transitions between eigenstates $|m\rangle$, $|n\rangle$ of $H$ with energy difference $\omega_{nm} = E_n - E_m$
(see App.~\ref{app:ME} for a derivation).
For the sake of notational simplicity we state the master equation for a single coupling term~\eqref{HI}. Multiple coupling terms lead to additional contributions of the same form.

The functions $\chi(\omega)$ and $\xi(\omega)$ in Eq.~\eqref{ME} follow from the environment spectral function
\begin{equation}
 \gamma(\omega) = 2 \pi \sum_\nu \lambda_\nu^2 \delta(\omega - \omega_\nu)
\end{equation}
and its analytical continuation $\Gamma(\omega)$ into the upper half plane,
with $\gamma(\omega) = \mp\Im \Gamma(\pm\omega + \rmi 0^+)$.
For a thermal environment with inverse temperature $\beta = 1 / T$
we get
\begin{equation}\label{chi}
\chi(\omega) = \begin{cases} \gamma(\omega) [n(\omega, T) + 1] \quad & \text{ if } \omega > 0 \;, \\[0.25ex]
 \gamma(-\omega) n(-\omega, T) & \text{ if } \omega < 0 
\end{cases}
\end{equation}
and
\begin{equation}\label{xi}
\xi(\omega) = \begin{cases} 
  \Re \Gamma(\omega + \rmi 0^+) [n(\omega, T) + 1]   \quad & \text{ if } \omega > 0 \;, \\[0.25ex]
 - \Re \Gamma(-\omega + \rmi 0^+) n(-\omega, T) & \text{ if } \omega < 0 \;,
\end{cases}
\end{equation}
with the Bose-Einstein distribution function
\begin{equation}
 n(\omega, T) = \frac{1}{\rme^{\beta \omega} - 1} \;.
\end{equation}
Note that in the zero temperature limit $n(\omega, T) \to 0$ such that the master equation~\eqref{ME} contains only dissipative terms for transitions $|n\rangle \to |m\rangle$ with positive energy $\omega_{nm} > 0$, i.e., dissipation correctly leads to energy decrease.
In particular, the problem of unphysical emission from the ground state encountered for the quantum optical master equations is resolved.

In the present work we assume an Ohmic spectral function $\gamma_c(\omega) = \gamma \omega / \omega_0$ for the cavity-environment coupling, and use $\gamma = 10^{-2} \omega_0$ in all numerical computations.
To reduce the number of free parameters we assume the same spectral function $\gamma_x^{(j)}(\omega) = \gamma_c(\omega)$ for the emitter-environment couplings.
The respective environment temperatures are also identical.

\subsection{Solution of the master equation}

As we show in App.\ \ref{app:ME}, the master equation\ \eqref{ME} 
splits into two equations of motion
\begin{eqnarray}\label{ME_nn}
 \frac{\rmd}{\rmd t} \rho_{n,n}(t) &=& \sum_{k \neq n} \chi(\omega_{kn}) S_{n,k} \rho_{k,k}(t) \nonumber\\
 && -\sum_{k \neq n} \chi(\omega_{nk}) S_{k,n} \rho_{n,n}(t) \,, \\\label{ME_mn}
 \frac{\rmd}{\rmd t} \rho_{m,n}(t) &=& -(Z_m + Z_n^*) \rho_{m,n}(t) \,, \quad (m \neq n)
\end{eqnarray}
for the matrix elements $\rho_{m,n}(t) = \langle m |{ \rho(t) }| n \rangle$ of the density operator.
In these equations, $S_{n,k} = |\langle n | S | k \rangle|^2$ and
\begin{equation}\label{Zn}
 Z_n = \frac{1}{2} \sum_{k \neq n} \big[ \chi(\omega_{nk}) + \rmi \xi(\omega_{nk}) \big] S_{k,n} + \rmi E_n \,.
\end{equation}
The general solution of Eq.\ \eqref{ME_mn} is
\begin{equation}\label{rho_mn}
 \rho_{m,n}(t) = \rme^{-(Z_m + Z_n^*) t} \rho_{m,n}(0) \,, \quad (m \neq n) \,.
\end{equation}
Because $\Re Z_n > 0$ for all $n$, the off-diagonal elements of $\rho(t)$ decay exponentially.
Hence, the stationary state fulfills
\begin{equation}\label{stat_diag}
 \rho_{m,n}^\infty \equiv \lim_{t \to \infty} \rho_{m,n}(t) = \rho_{n,n}^\infty \delta_{m,n} \,.
\end{equation}
The diagonal elements $\rho_{n,n}^\infty$ are determined by the stationary solution of the Pauli master equation\ \eqref{ME_nn}.
If the system is coupled to a thermal environment as in Eqs.~\eqref{chi} and\ \eqref{xi}, 
the stationary solution of Eq.\ \eqref{ME_nn} is the thermal state $\rho^\infty \propto \rme^{-\beta H}$ of the system corresponding to the temperature $T=1/\beta$ of the environment.

The emission spectrum and photon statistics 
can now be computed through a standard input-output formalism (see App.\ \ref{app:inout}),
which leads to the projected cavity-environment coupling operator
\begin{equation}\label{X}
 \dot{X}_- = -\rmi \sum_{m,n>m} (E_n - E_m) | m \rangle \langle m | X | n \rangle \langle n |
\end{equation}
describing the emission.
The correlation functions of $\dot{X}_-$ and $\dot{X}_+ = (\dot{X}_-)^\dagger$ characterize the properties of the emitted light. The emission spectrum of the cavity is
\begin{equation}\label{spectrum}
 S(\omega) = \lim_{t \to \infty} \frac{\gamma_c(\omega)}{\pi} \Re \int_0^\infty \rme^{-\rmi \omega \tau} \langle \dot{X}_+(t + \tau) \dot{X}_-(t) \rangle \rmd\tau \,,
\end{equation}
and the second-order Glauber function \cite{Gla63} reads
\begin{equation}\label{glauber}
 g^{(2)}(\tau) = \lim_{t \to \infty} \frac{\langle \dot{X}_+(t) \dot{X}_+(t + \tau) \dot{X}_-(t + \tau) \dot{X}_-(t) \rangle}{\langle \dot{X}_+(t) \dot{X}_-(t) \rangle^2} \,.
\end{equation}
Note that evaluation of Eqs.~\eqref{spectrum},~\eqref{glauber} requires diagonalization of the Hamiltonian $H$.

Because the stationary state $\rho^\infty$ from Eq.~\eqref{stat_diag} is diagonal in the eigenbasis $|n\rangle$ of $H$, we can evaluate the $\tau$-integration in Eq.\ \eqref{spectrum} analytically as
\begin{eqnarray}\label{S_SumLorentz}
 S(\omega) &=& \frac{\gamma_c(\omega)}{\pi} \sum_{m < n} |\langle m | \dot{X}_- | n \rangle|^2 \rho_{n,n}^\infty \\
 && \times \frac{\Re (Z_m+Z_n)}{ [\omega- \Im (Z_n - Z_m)]^2 + [\Re (Z_m+Z_n)]^2 } \,. \nonumber
\end{eqnarray}
The emission spectrum $S(\omega)$ is the sum of Lorentz peaks 
with width $\Re (Z_n + Z_m)$ at the respective transition energies $\Im (Z_n - Z_m)$, which according to Eq.~\eqref{Zn}
are shifted relative to the transition energies $E_n-E_m$ of the closed system by a Lamb shift that results from coupling to the environment.

\subsection{Quantum optical master equation}

It is instructive to compare the master equation~\eqref{ME} to the quantum optical master equation~\cite{Car99}
\begin{eqnarray}\label{ME_QO}
 \frac{\rmd}{\rmd t} \rho(t) &=& -\rmi [H, \rho(t)] - \rmi \sum_\pm \frac{\xi_\pm}{2} [S_\pm^\dagger S_\pm, \rho(t)] \\
 && + \sum_\pm \frac{\chi_\pm}{2} \Big( \big[ S_\pm \rho(t), S_\pm^\dagger \big] + \big[ S_\pm, \rho(t) S_\pm^\dagger \big] \Big) \;, \nonumber
\end{eqnarray}
which is obtained by replacing the projected operators $S_\omega$ with the `bare' operators $S_\pm = \sum_{\omega \gtrless 0} S_\omega$,
and by assuming $\chi(\pm\omega) \approx \chi_\pm$, $\xi(\pm\omega) \approx \xi_\pm$
in the vicinity of a typical transition energy $\omega$.
Note that $S_+ = -\rmi X_0 a$ for the cavity-environment coupling and $S_+ = -\rmi \sigma_-^{(j)}$ for the emitter-environment coupling.
Evidently this approximation can be valid only for weak light-matter coupling $g, g' \ll \omega_c, \omega_x$, 
when the dressing of emitter states by cavity photons can be neglected.
Because the quantum optical master equation
does not distinguish between energy-increasing and energy-decreasing transitions, which are equally contained in the unprojected operator $S$ because of hermiticity, it can lead to unphysical predictions 
such as emission out of the ground state.
Furthermore, because failure to observe the above distinction is tantamount to a high-temperature approximation, one will expect that the quantum optical master equation 
fails at the prediction of non-thermal photon statistics at low temperatures.
Therefore, we use the more general master equation~\eqref{ME}.

\section{\label{sec:spectra}The emitted light}
The first characterization of the light generated in the cavity is provided by the emission spectrum.
Because the emission spectrum depends on the (Lamb-shifted) energy spectrum of the Dicke Hamiltonian we start with a discussion of the eigenvalues of $H$ for few emitters, before we turn to the actual function $S(\omega)$ obtained from numerical solution of the master equation\ \eqref{ME}.

\subsection{Energy spectrum of the Hamiltonian}

To construct the energy spectrum of $H$ we notice first that the eigenstates of $N$ uncoupled two-level emitters 
can be classified as angular momentum eigenstates 
with total angular momentum $J = N/2, N/2-1, \ldots \geq 0$.
Since $H$ commutes with the total angular momentum operator,
states with different $J$ do not mix even at finite coupling $g, g' \ne 0$.
For fixed $J$, the $J_z$ quantum number $M$ can assume the values $M = -J, -J+1, \dots, J$,
and a corresponding emitter eigenstate has energy $(M + N/2) \omega_x$. 
Note that for $N \ge 3$ the classification in terms of $J$, $M$ is not exhaustive,
since different emitter states can have identical values. However, these states give the same contribution to the emission spectrum.
The cavity photon eigenstates are Fock states $|n\rangle$ with energies $n \omega_c$.

\begin{figure}
 \includegraphics{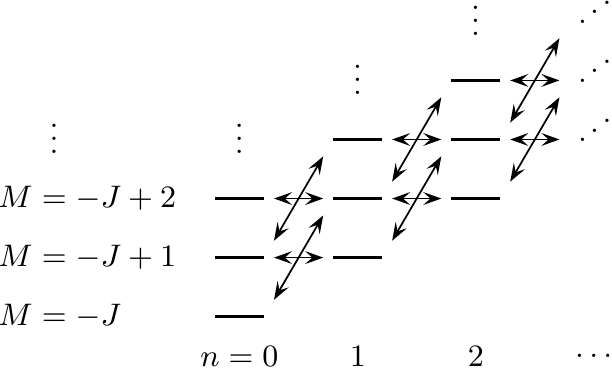}
 \caption{\label{fig:ladder}Schematic energy level pattern for the construction of the spectrum of the Dicke model.
  Horizontal arrows depict the co-rotating interaction terms in Eq.\ \eqref{HS} (coupling constant $g$), diagonal arrows depict the counter-rotating terms ($g'$).}
\end{figure}

\begin{figure}
 \includegraphics[width=0.95\linewidth]{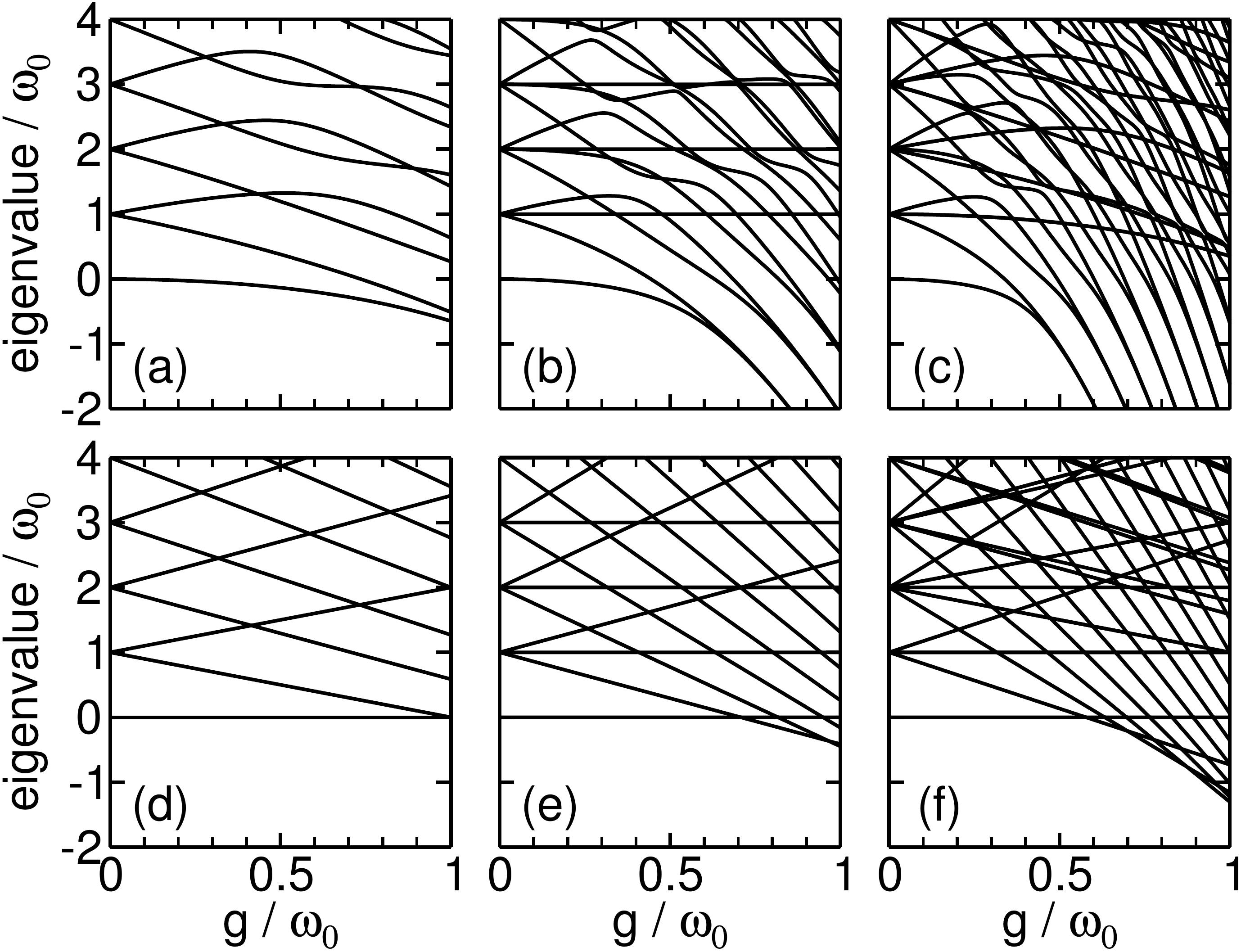}
 \caption{\label{fig:E}Eigenvalues of $H$ in Eq.\ \eqref{HS} for $g'=g$ (panels a-c) and $g'=0$ (panels d-f) as functions of the coupling strength $g$.
  Panels (a,b) show the results for one emitter, whereas the number of emitters is $N = 2$ in panels (b,e) and $N = 3$ in panels (c,f).}
\end{figure}

For given $J$ we can arrange the eigenstates of the uncoupled emitter-cavity system as the rungs of a ladder diagram as in Fig.\ \ref{fig:ladder}.
Working at resonance $\omega_c = \omega_x = \omega_0$,
the energy $(M + N/2 + n) \omega_0$ of each state is given by the total number of emitter and cavity excitations.
The co-rotating light-matter interaction terms in $H$ preserve the number of excitations and connect states at the same energy level (horizontal arrows in Fig.\ \ref{fig:ladder}).
The counter-rotating terms change the number of excitations by two (diagonal arrows in Fig.\ \ref{fig:ladder}).
This simple scheme explains many properties of the energy spectra of $H$ shown in Fig.\ \ref{fig:E}.

For $N = 1$ it is $J=1/2$,
and we recover the Jaynes-Cummings ladder~\cite{JC63} for $g'=0$.
The lowest level ($M=-1/2$, $n=0$) does not couple to any other state and hence leads to the $g$-independent eigenvalue zero of $H$ [see Fig.\ \ref{fig:E}(d)].
Every other level consists of two ladder rungs.
They lead to the eigenvalues $n \omega_0 \pm \sqrt{n} g$, for $n \geq 1$.
For $g'=g$ corrections arise from coupling between states at different height in the ladder but the energy level pattern remains discernible [Fig.\ \ref{fig:E}(a)].

For $N = 2$  [see Fig.\ \ref{fig:E}(b,e)] we have either triplet ($J = 1$) or singlet ($J = 0$) emitter states.
For $g'=0$, the triplet states lead to the eigenvalue zero ($n=0$), the two eigenvalues $\omega_0 \pm \sqrt{2} g$ ($n=1$), and the three eigenvalues $n \omega_0$, $n \omega_0 \pm \sqrt{2} \sqrt{2 n - 1} g$ for $n \geq 2$.
The singlet states do not couple with each other and lead to the $g$-independent eigenvalues $(n+1) \omega_0$ for $n \geq 0$.
It follows that the eigenvalues $n \omega_0$ for $n \geq 2$ are twofold degenerate (one triplet, one singlet state).
This degeneracy is lifted for $g'=g$, but the energies of the singlet states remain fixed.

For $N = 3$ we have quadruplet ($J = 3/2$) and doublet ($J = 1/2$) emitter states.
The ladder scheme for the doublet is equal to that for $N = 1$ and hence leads to the same energy spectrum, apart from the fact that all energies are shifted up by $\omega_0$ when going from $N=1$ to $N=3$.
Notice that the doublet states are two-fold degenerate, because the angular momentum classification of the emitter states is not unique in this case.
The quadruplet states lead to one (starting at zero for $g = 0$), two (at $\omega_0$), three (at $2 \omega_0$), and four (at $n \omega_0$ with $n \geq 3$) additional eigenvalues in Figs.\ \ref{fig:E}(f).
Because of the close vicinity of many states in the energy spectrum the corrections resulting from the counter-rotating terms for $g'=g$ are large.
This trend continues if $N$ is increased further.


\subsection{The emission spectrum}

\begin{figure}
 \includegraphics[width=0.49\linewidth]{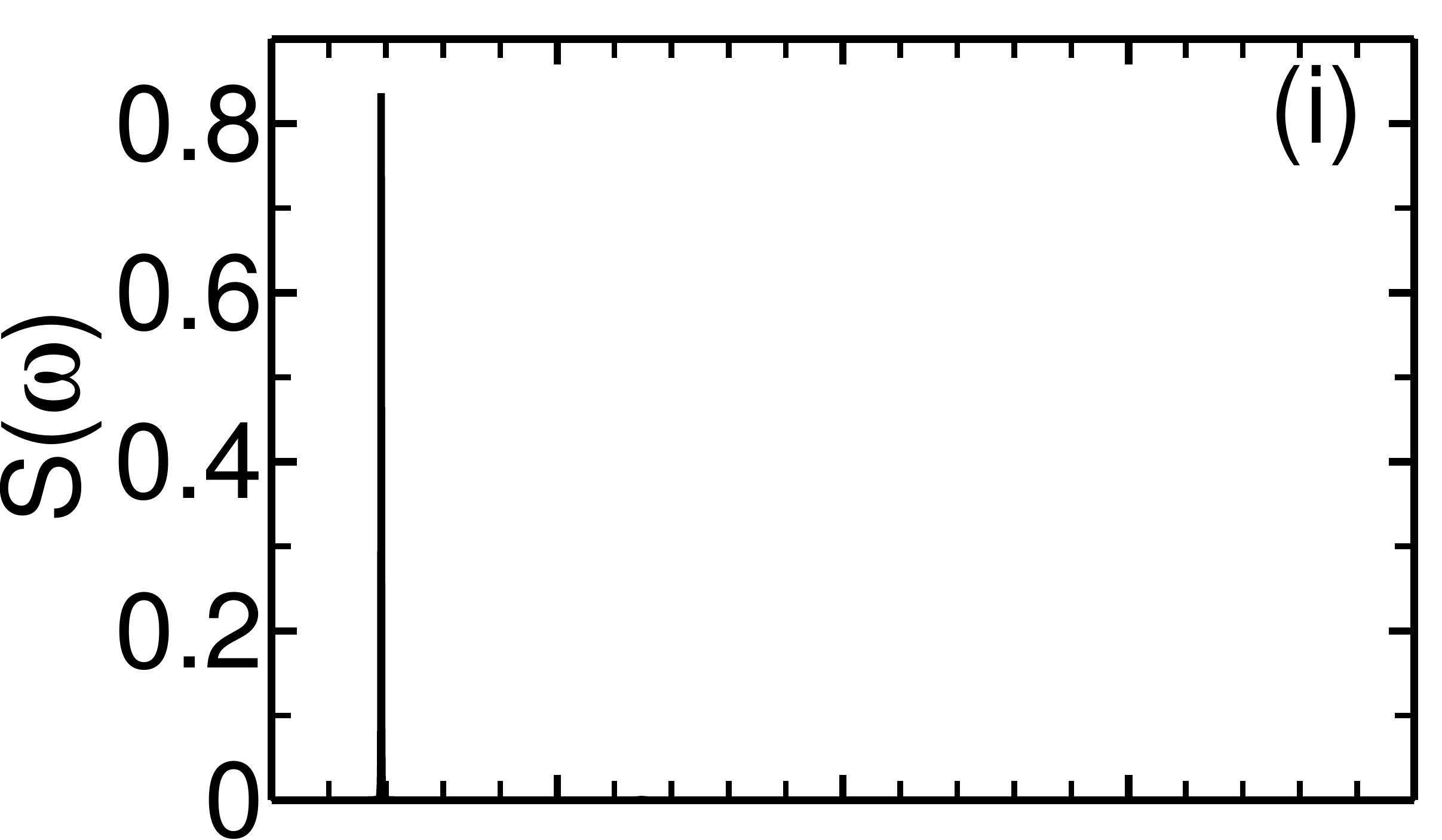}
 \includegraphics[width=0.49\linewidth]{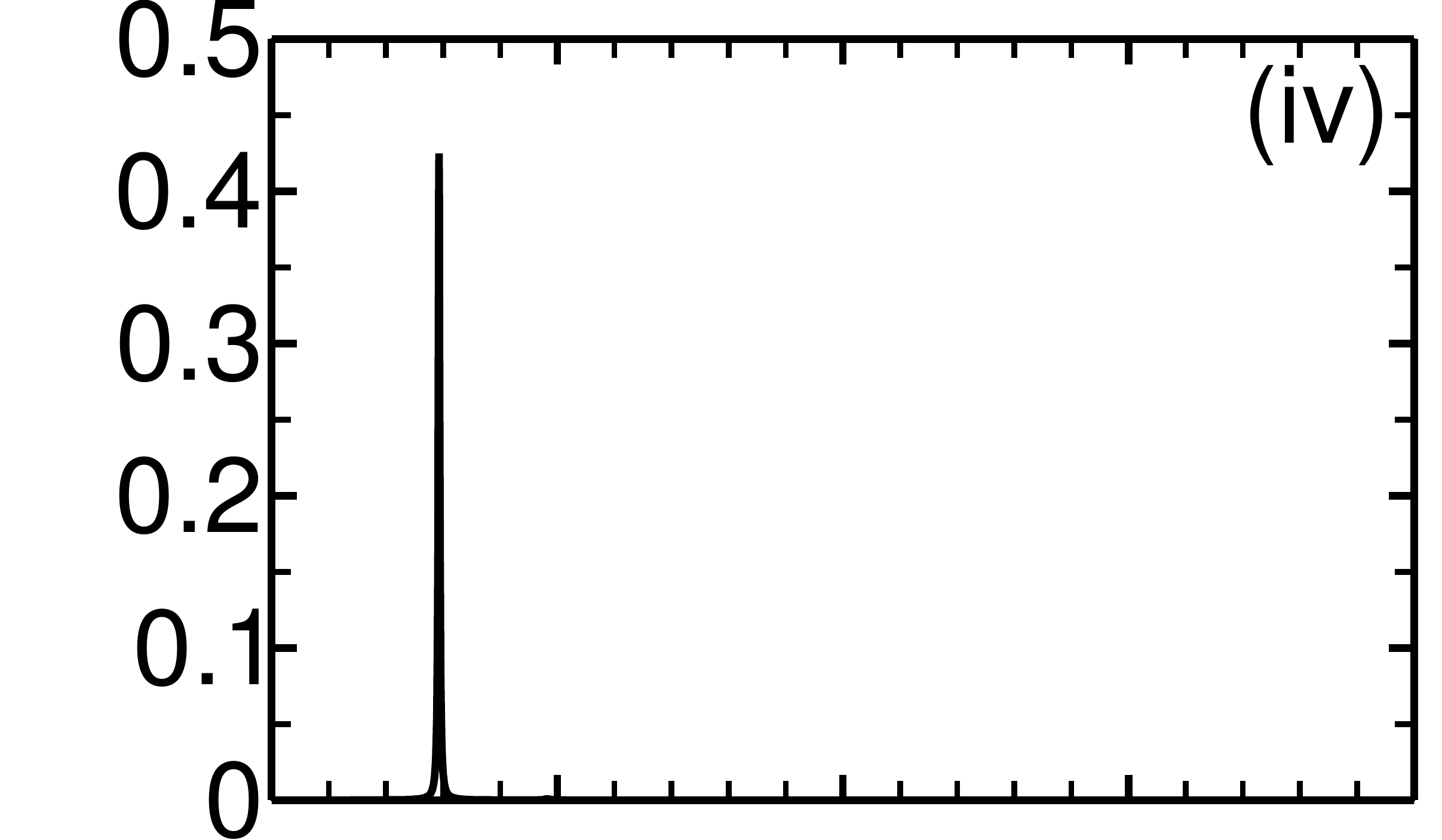}\\
 \includegraphics[width=0.49\linewidth]{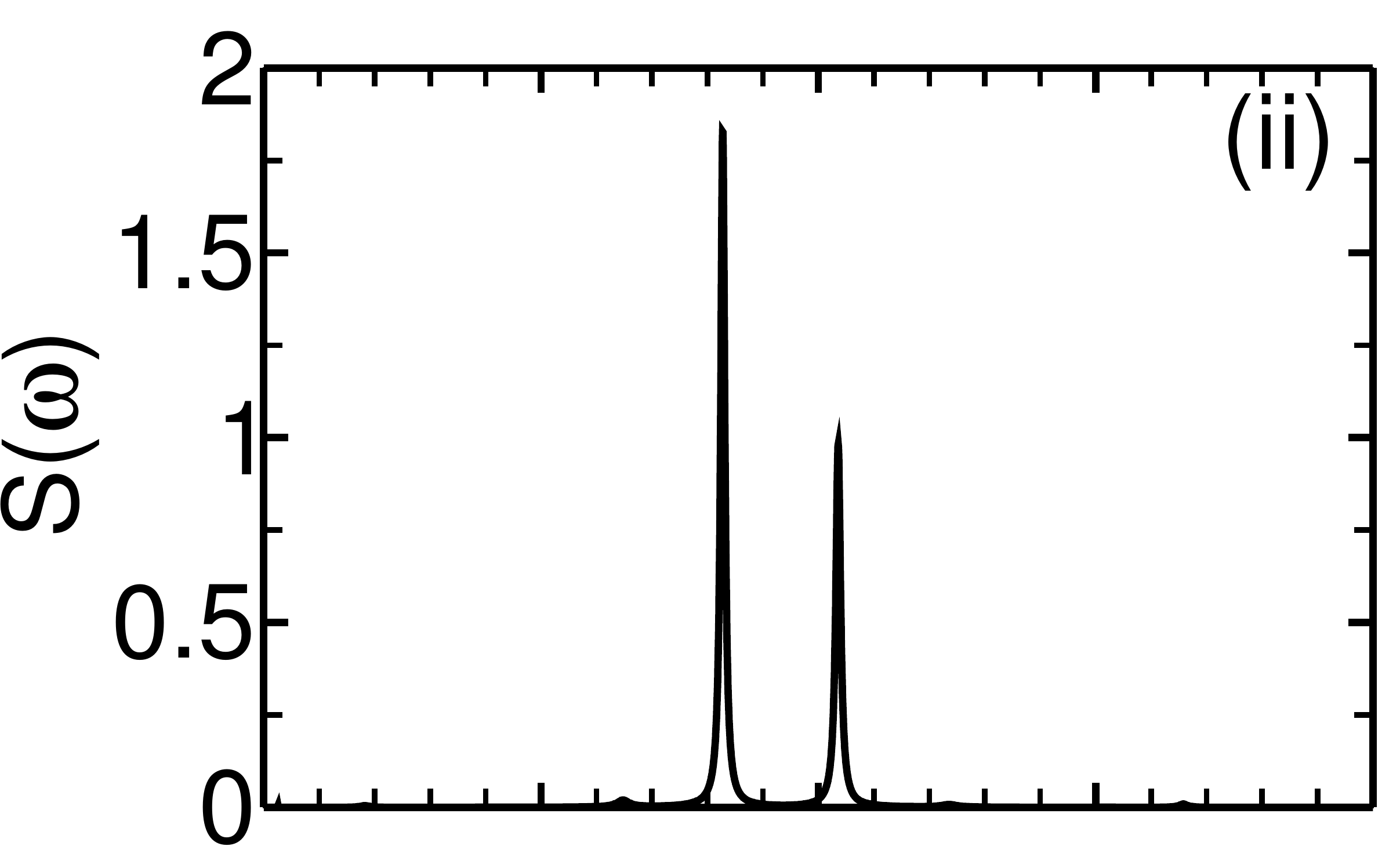}
 \includegraphics[width=0.49\linewidth]{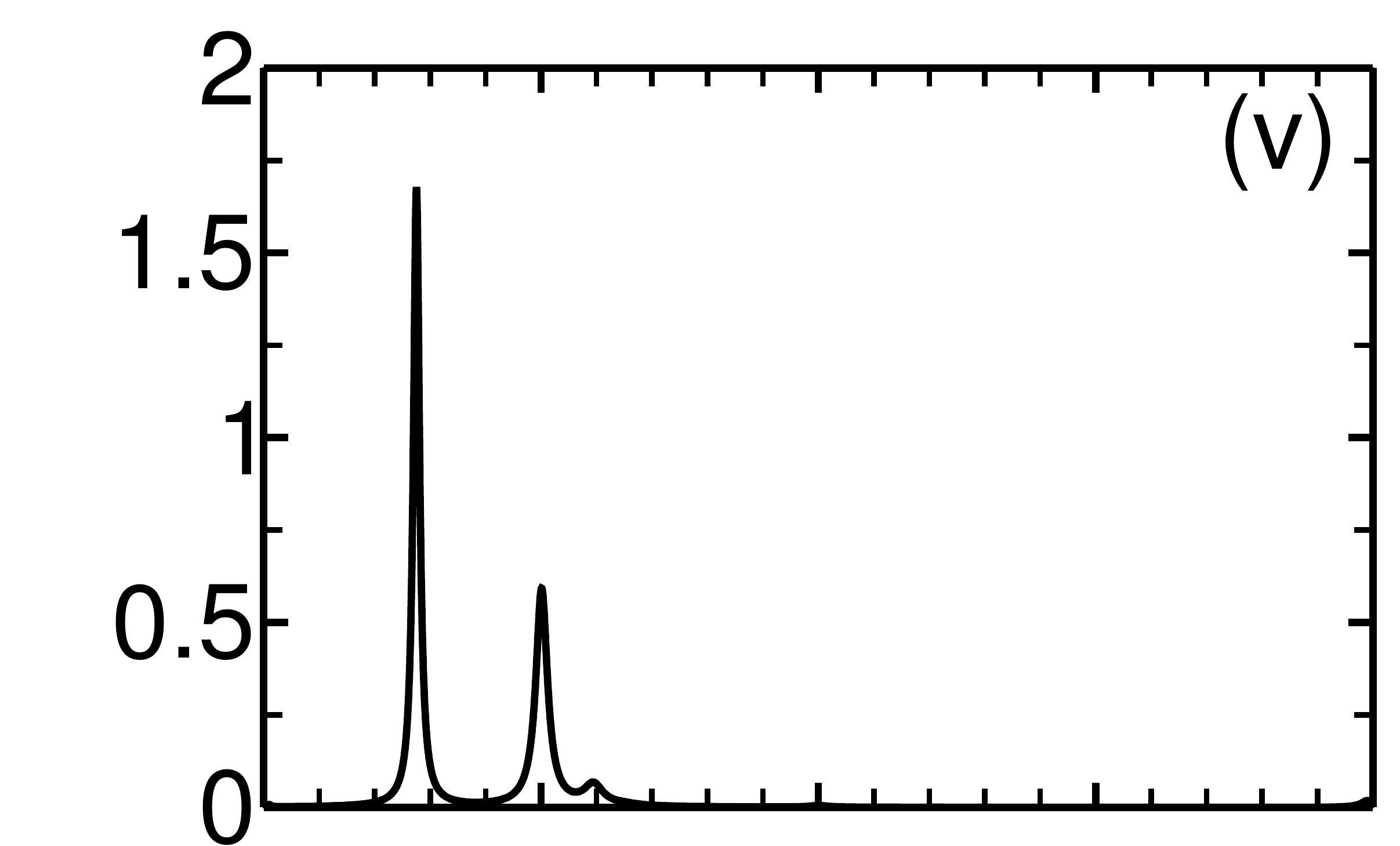}\\
 \includegraphics[width=0.49\linewidth]{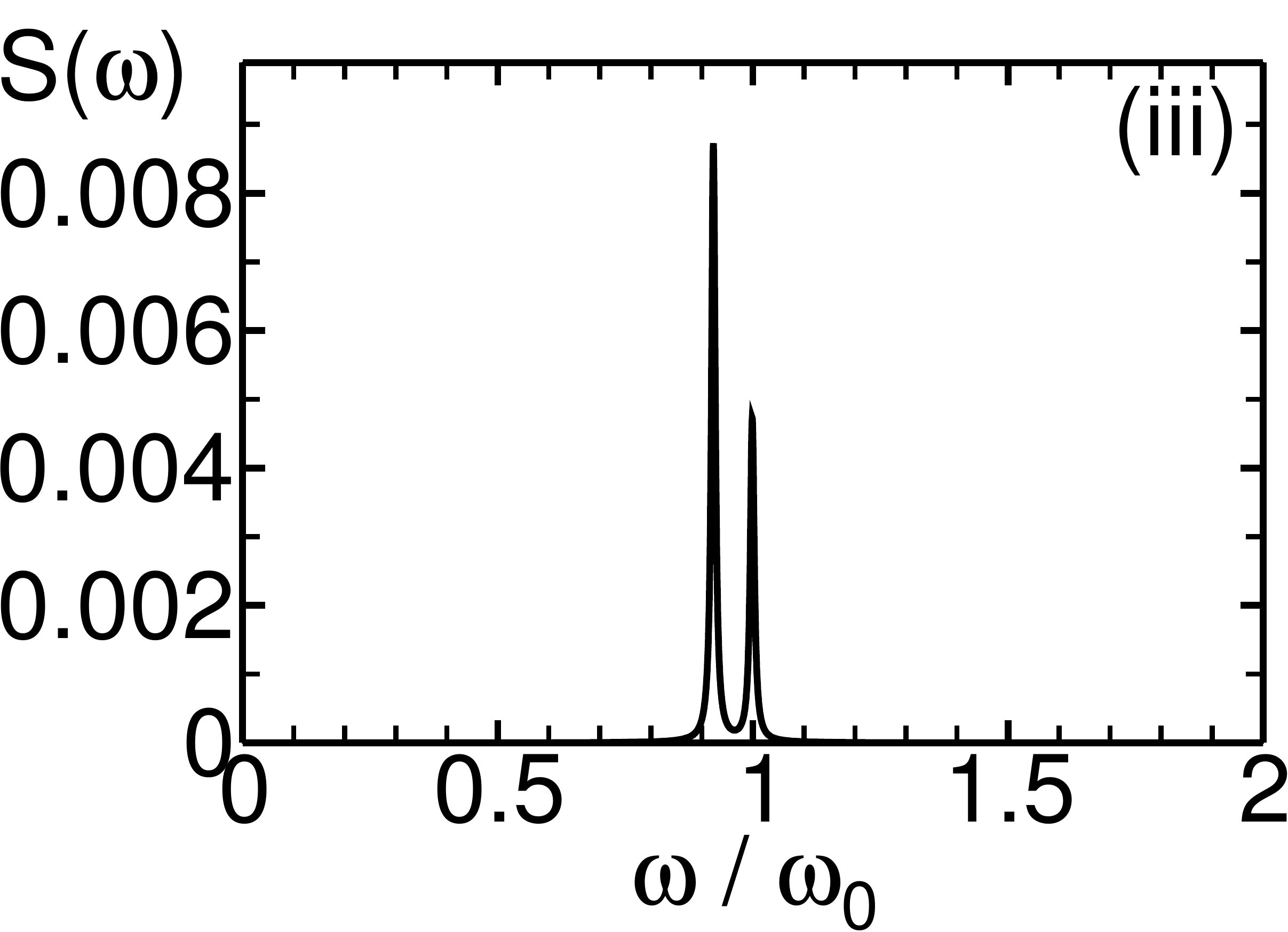}
 \includegraphics[width=0.49\linewidth]{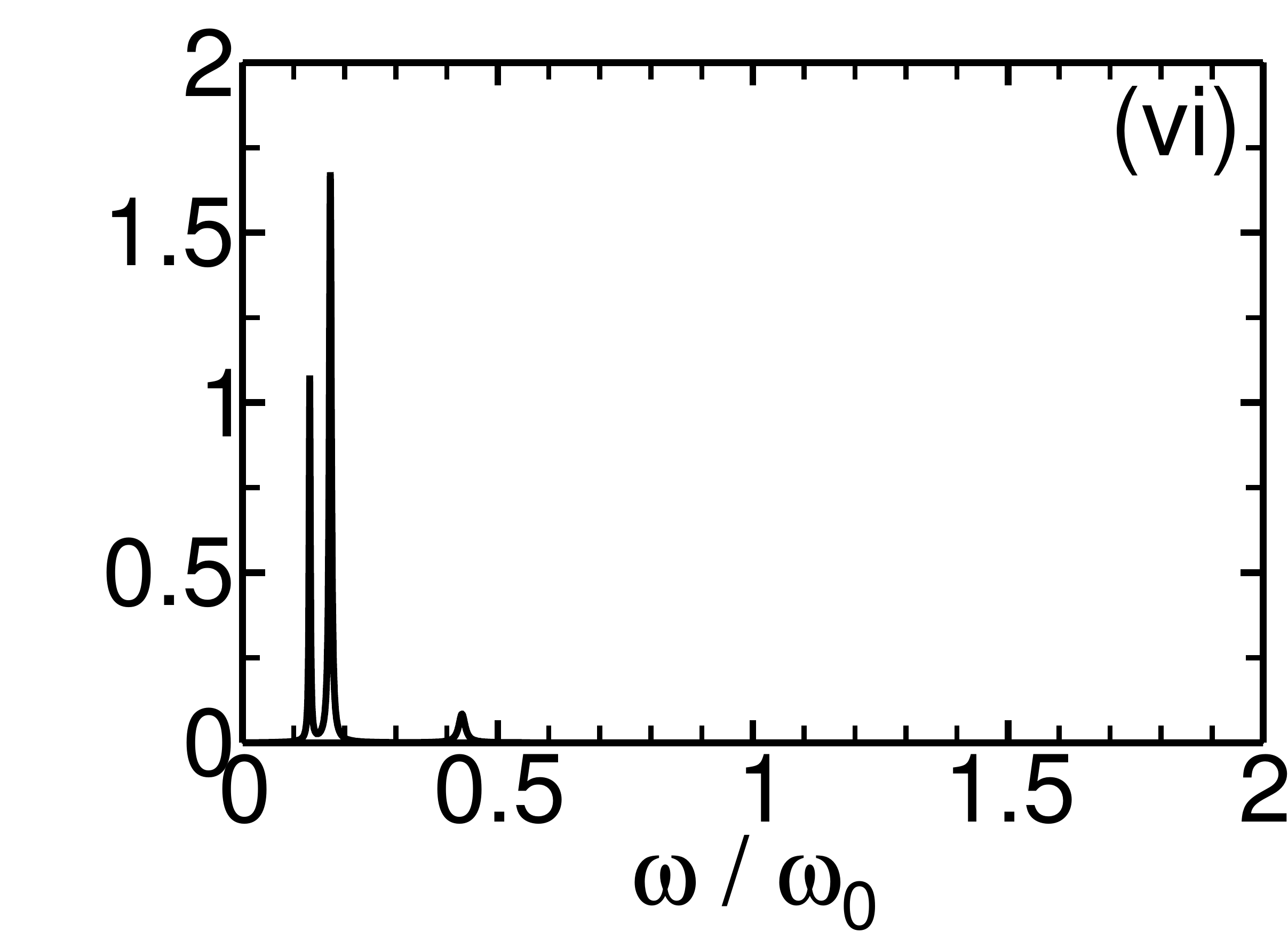}
 \caption{\label{fig:spectra}Emission spectra $S(\omega)$ for $N=2$ emitters.
  Panels (i-iii) show results for $g'=g$, panels (iv-vi) for $g'=0$.
  The emitter-cavity coupling strength and the environment temperature are $g = 0.5 \omega_0$, $T = 0.07 \omega_0$ in panels (i, iv), $g = 0.7 \omega_0$, $T = 0.23 \omega_0$ in panels (ii,v), and $g = 0.8 \omega_0$, $T = 0.1 \omega_0$ in panels (iii,vi).}
\end{figure}

In Fig.\ \ref{fig:spectra} we show the emission spectrum $S(\omega)$ for $N=2$ emitters at different coupling strength $g$ and environment temperature $T$. 
These data, as well as those for the Glauber function $g^{(2)}(t)$ shown later, have been computed  with a maximal number of $10^2$ cavity photons in the numerical diagonalization of $H$, which is sufficient for the given parameter combinations.

For low temperatures $T \ll \omega_0$, only the first possible transition into the ground state contributes to the emission spectrum.
It leads to the single peak in panels (i) and (iii) of Fig.\ \ref{fig:spectra}.
With increasing temperature transitions involving higher excited states begin to contribute.
For example the two peaks in  panels (ii), (iv) correspond to the transition
from the 2nd to the 1st excited state and from the 3rd excited to the ground state.
As could be deduced already from panels (b), (e) in Fig.\ \ref{fig:E},
the transitions tend to have smaller energies in the TC limit than in the Dicke limit, which leads to the red shift of the emission peaks in panel (v) relative to those in panel (ii). 
However, at not too strong coupling the low-lying states 
still have comparable energies, and the emission spectra look similar.
The situation changes at ultrastrong coupling when the co-rotating and counter-rotating terms are of equal magnitude (panels (iii), (vi)). 
In addition to the markedly different peak energies the peak height has now decreased by two orders of magnitude in the Dicke limit, but not in the TC limit.

\begin{figure}
 \includegraphics[width=0.49\linewidth]{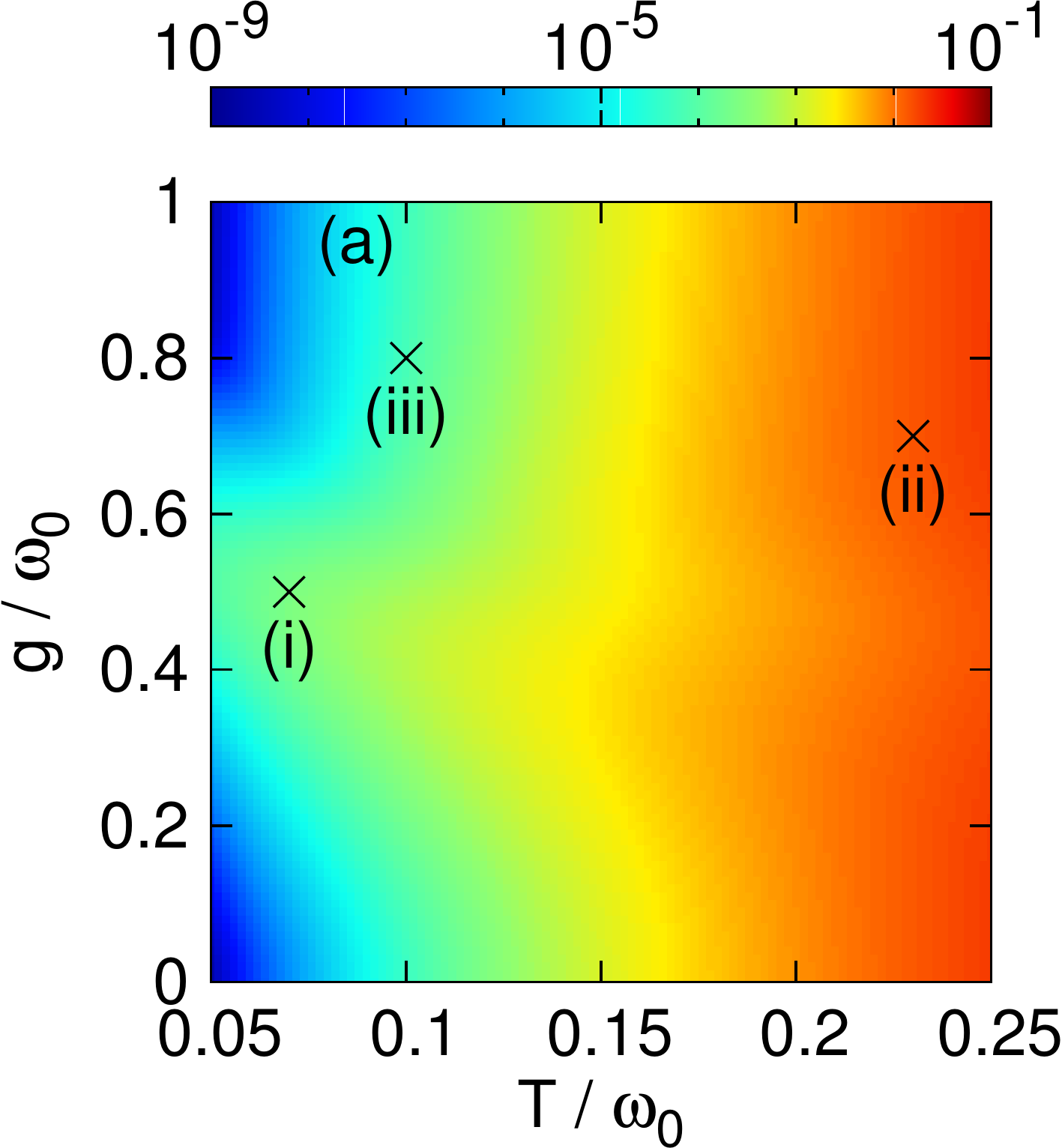}
 \includegraphics[width=0.49\linewidth]{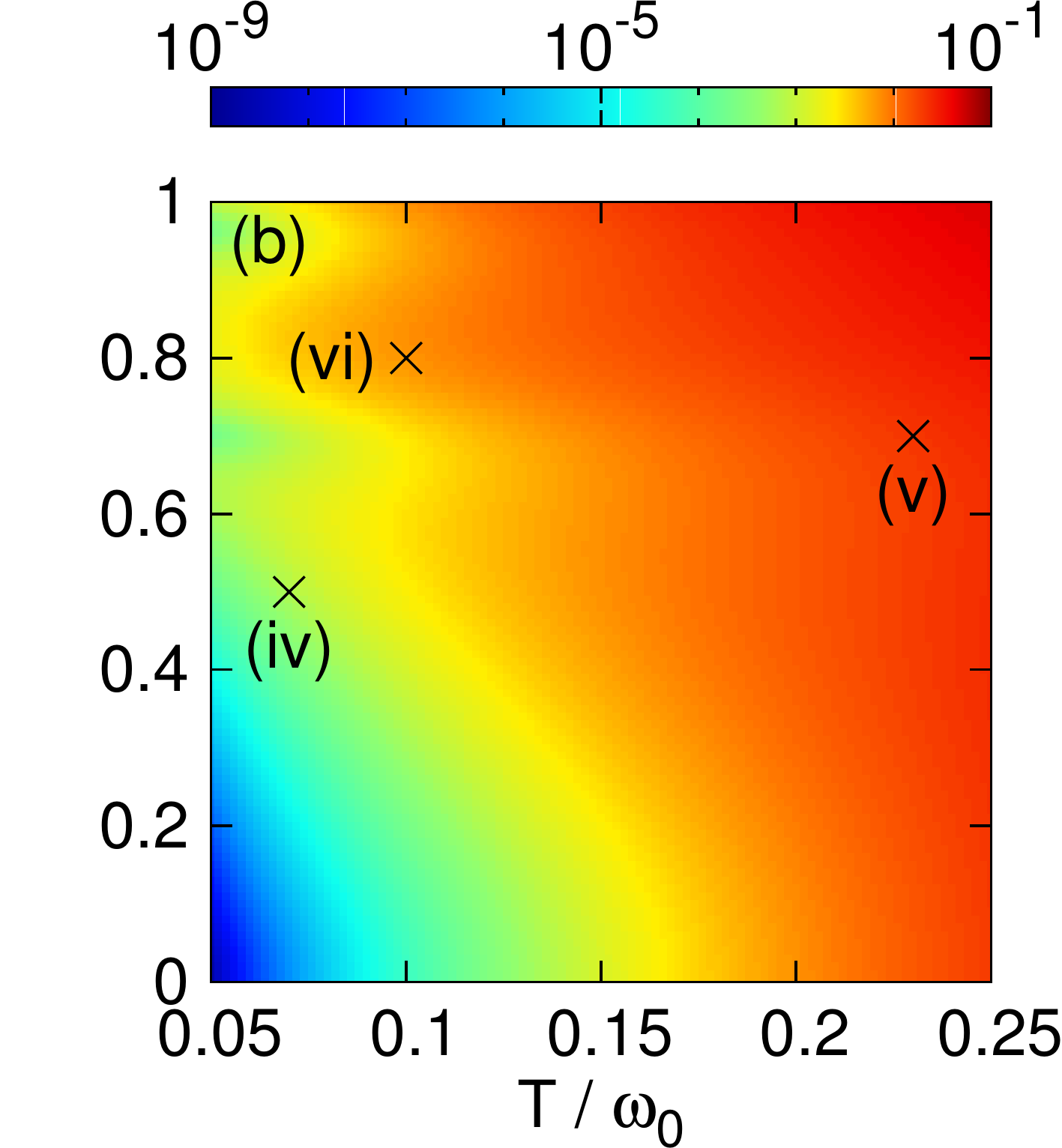}
 \caption{\label{fig:intSpectra}(Color online) Expectation value $\langle \dot{X}_+ \dot{X}_- \rangle$ for $N=2$ emitters, as a function of temperature $T$ and coupling strength $g$.
  Panel (a) shows the result for $g'=g$, panel (b) for $g'=0$.
Crosses mark the parameters used in Fig.~\ref{fig:spectra}.
}
\end{figure}

The decrease of peak height can be recognized in the $\omega$-integrated emission spectrum
\begin{equation}
\int_{-\infty}^\infty \frac{S(\omega)}{\gamma_c(\omega)} \rmd\omega  = \langle \dot{X}_+ \dot{X}_- \rangle 
\end{equation}
shown in Fig.~\ref{fig:intSpectra}.
The equality with the given expectation value follows directly from Eq.~\eqref{S_SumLorentz}.
Only in the Dicke limit, but not in the TC limit, 
the total emission becomes small again at ultrastrong coupling and low temperatures.
Still, one sees that both plots agree nicely for not too strong coupling ($g/\omega \lesssim 0.5$).
This observation sets the upper limit of the coupling strength (here, for $N=2$ emitters) below which the presence or absence of counter-rotating interaction terms does not affect the light emission significantly.
We will find the same behavior for the Glauber function.

\section{\label{sec:statistics}Nonclassical light}
A basic decision on the possible generation of nonclassical light is possible with the Glauber function $g^{(2)}(0)$ at zero time delay.
For $g^{(2)}(0) = 1$ the emitted photons have a Poissonian distribution,
while $g^{(2)}(0) > 1$ indicates super-Poissonian statistics.
Thermal light has $g^{(2)}(0) = 2$.
By contrast, $g^{(2)}(0) < 1$ indicates nonclassical light with sub-Poissonian photon statistics.
Further information on photon (anti-)bunching
is provided by the full time-dependent function $g^{(2)}(t)$.

\subsection{Photon statistics for one emitter}

\begin{figure}
 \includegraphics[width=0.49\linewidth]{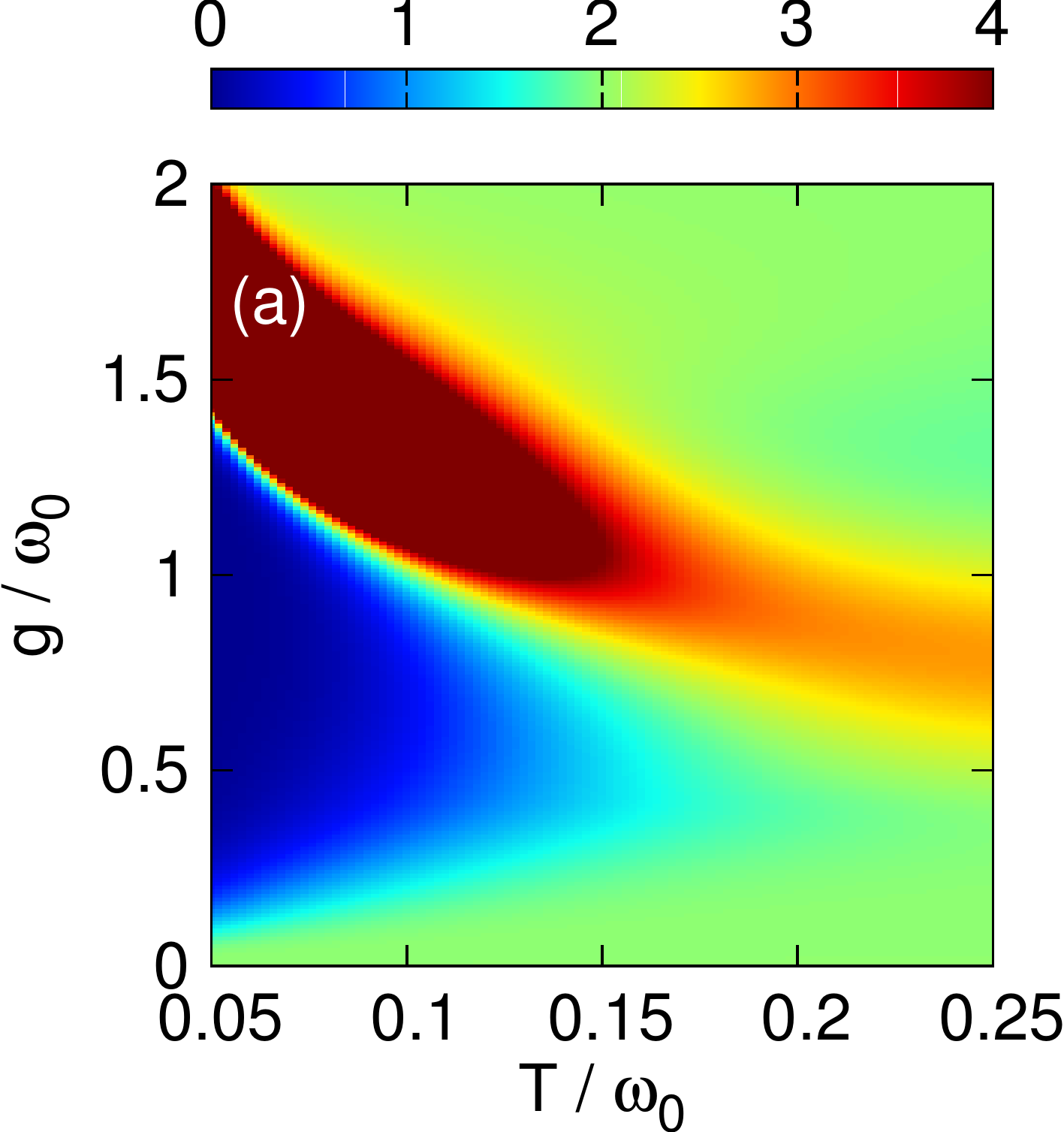}
 \includegraphics[width=0.49\linewidth]{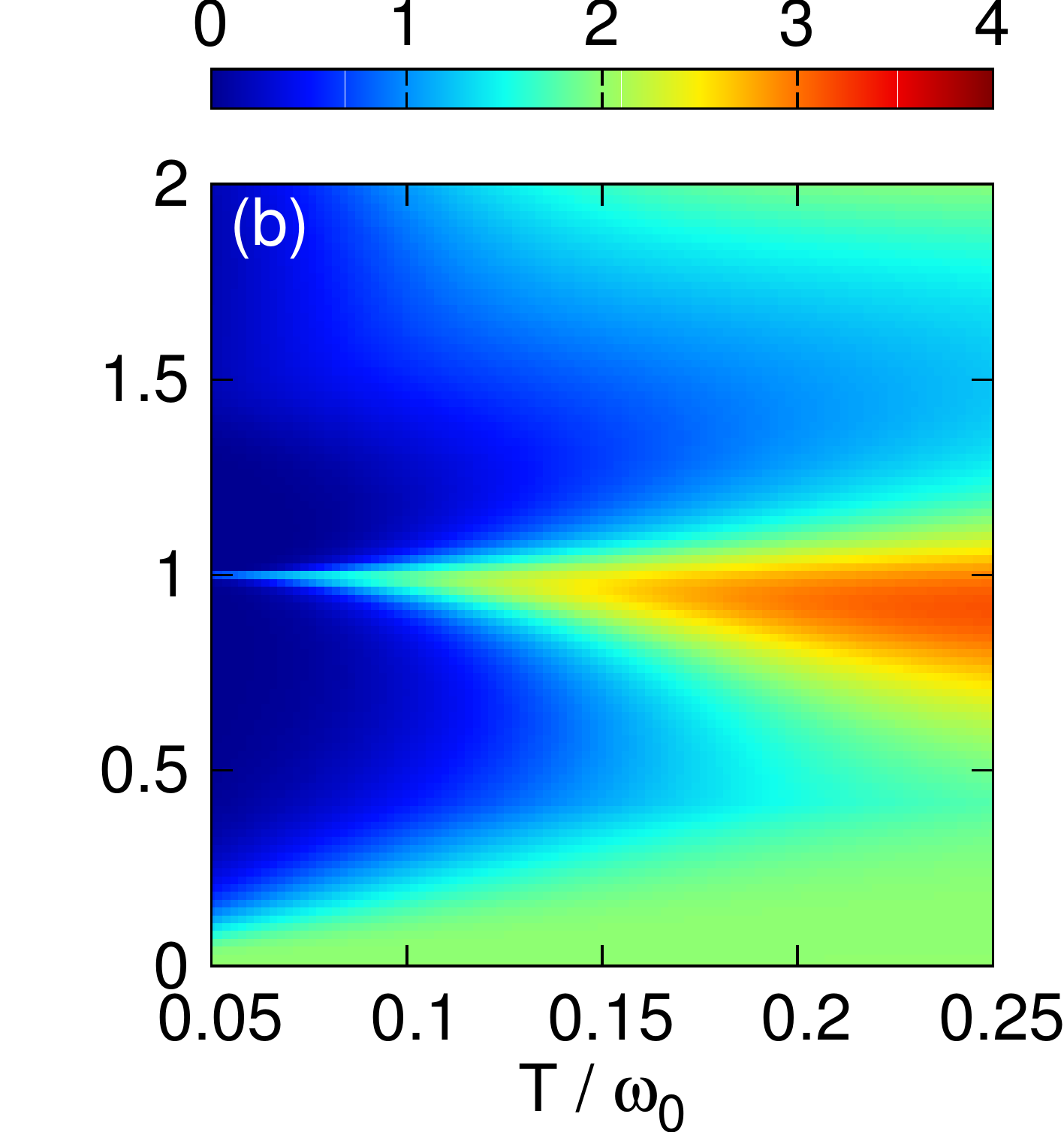}
 \caption{\label{fig:g2_chart_ext}(Color online) Glauber function $g^{(2)}(0)$ at zero time-delay for one emitter ($N=1$), as a function of temperature $T$ and coupling strength $g$.
  Panel (a) shows the result for $g'=g$, panel (b) for $g'=0$.
  Note that all values $g^{(2)}(0) \ge 4$ are assigned the same (dark-red) color in the density plots.}
\end{figure}

The Glauber function $g^{(2)}(0)$ for one emitter ($N = 1$) is shown in Fig.\ \ref{fig:g2_chart_ext}.
Two distinct regions can be identified in the Dicke limit in panel (a) (where $g'=g$).
A triangular region with $g^{(2)}(0) < 1$,
which stretches out along the vertical axis,
indicates the emission of nonclassical light with sub-Poissonian photon statistics
at low temperatures and moderate-to-strong light-matter coupling.
It lies below an elongated region with strongly super-Poissonian photon statistics ($g^{(2)}(0) \gg 2$) at larger coupling,
which extends diagonally towards higher temperatures.
Both regions are embedded in the background of thermal light with $g^{(2)}(0) \approx 2$.
The situation is distinctly different in the TC limit ($g'=0$) in panel (b),
where the super-Poissonian region is pushed back in favor of a second sub-Poissonian region that continues towards ultrastrong coupling.
Note, however, that the emission of nonclassical light in the first sub-Poissonian region is observed equally in both limits.

\begin{figure}
 \includegraphics[width=0.49\linewidth]{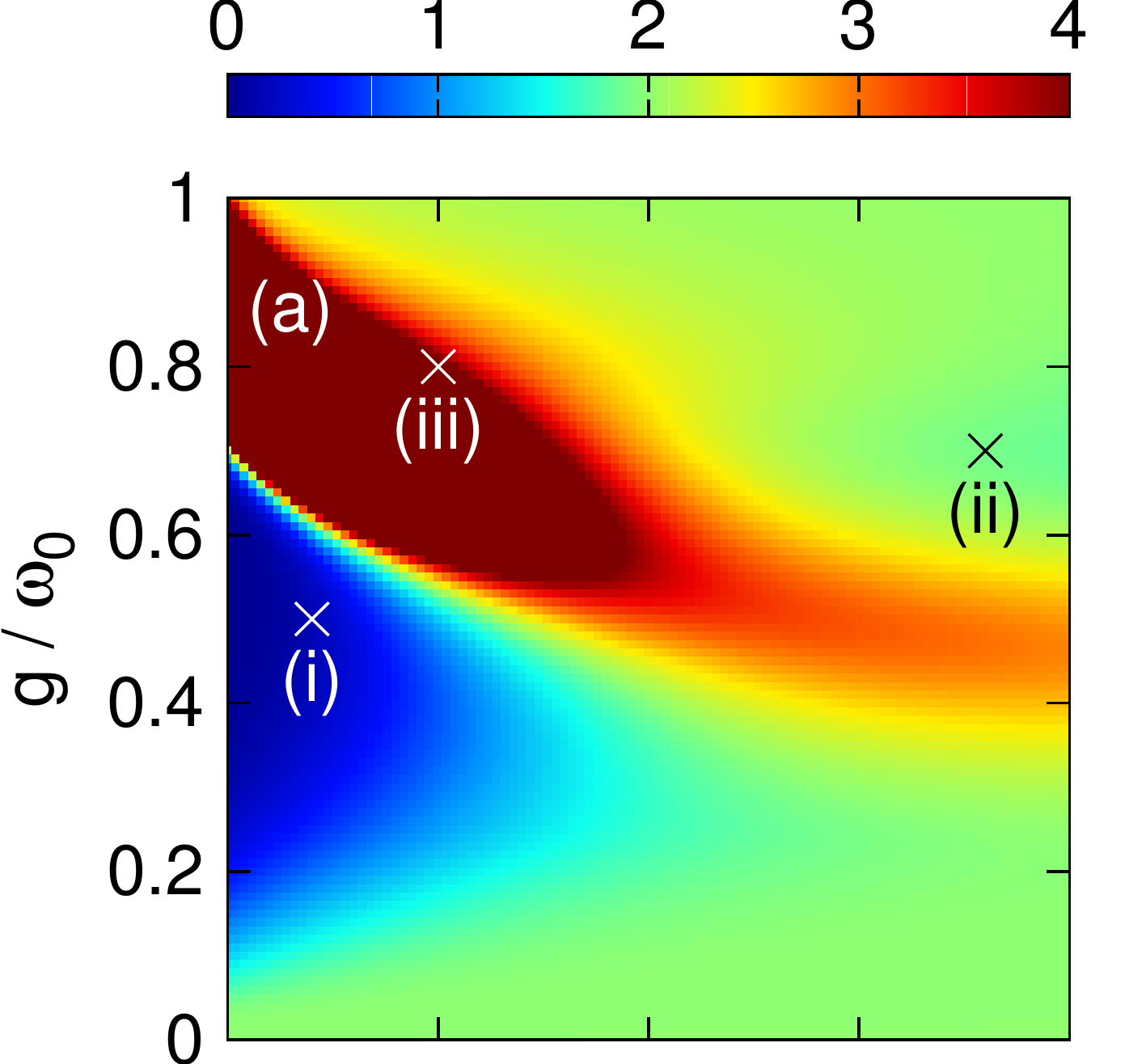}
 \includegraphics[width=0.49\linewidth]{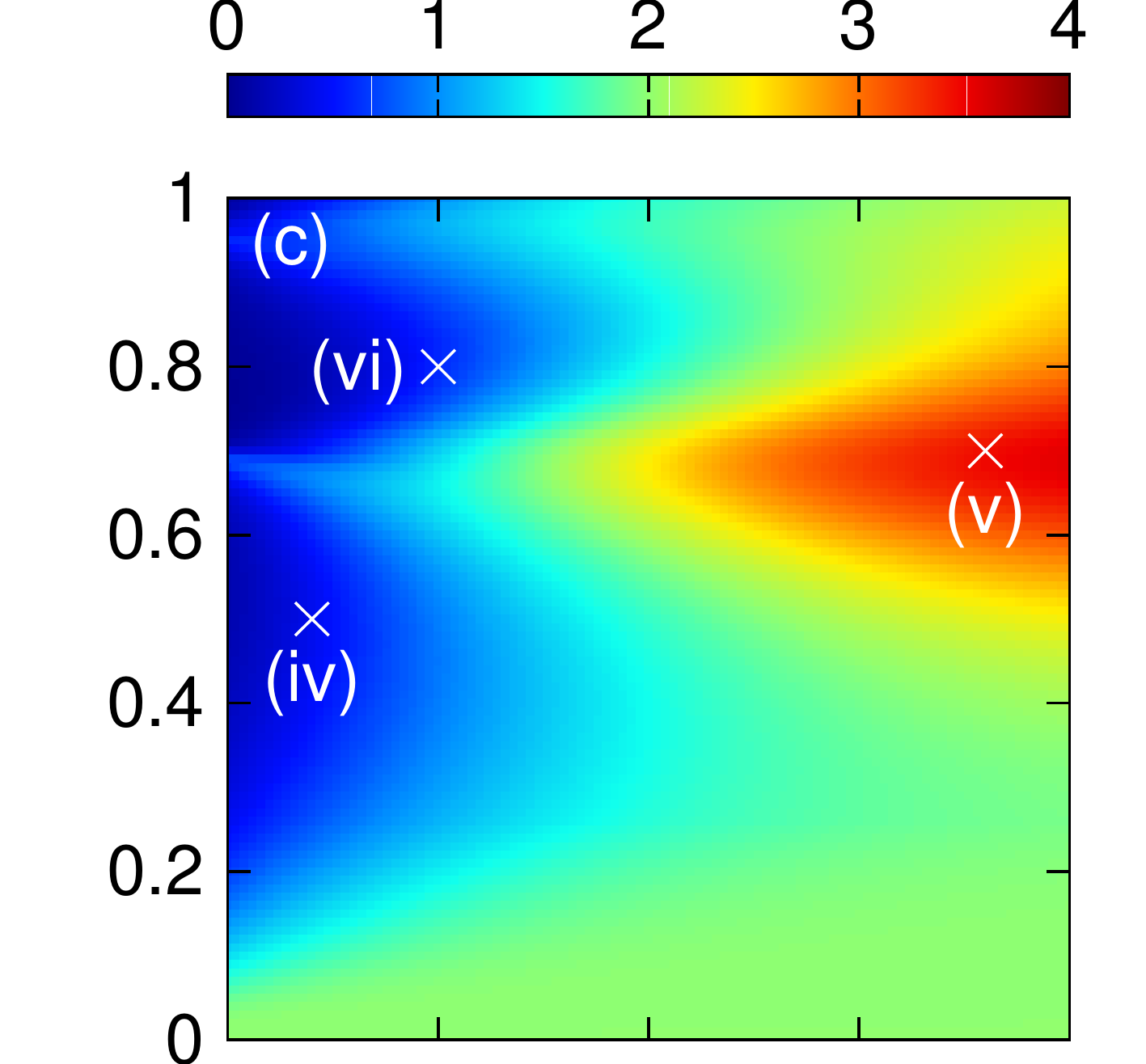}\\
 \includegraphics[width=0.49\linewidth]{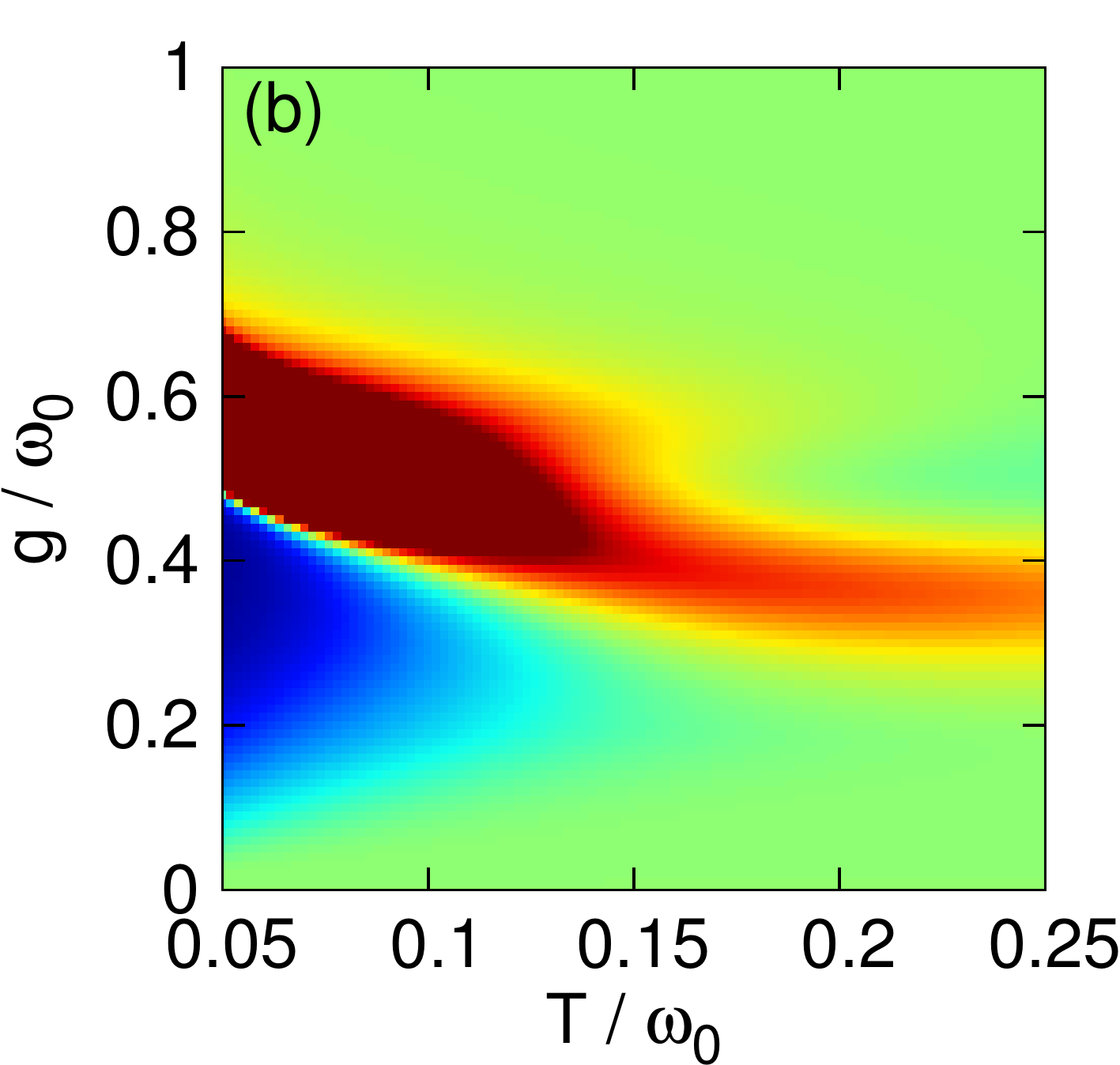}
 \includegraphics[width=0.49\linewidth]{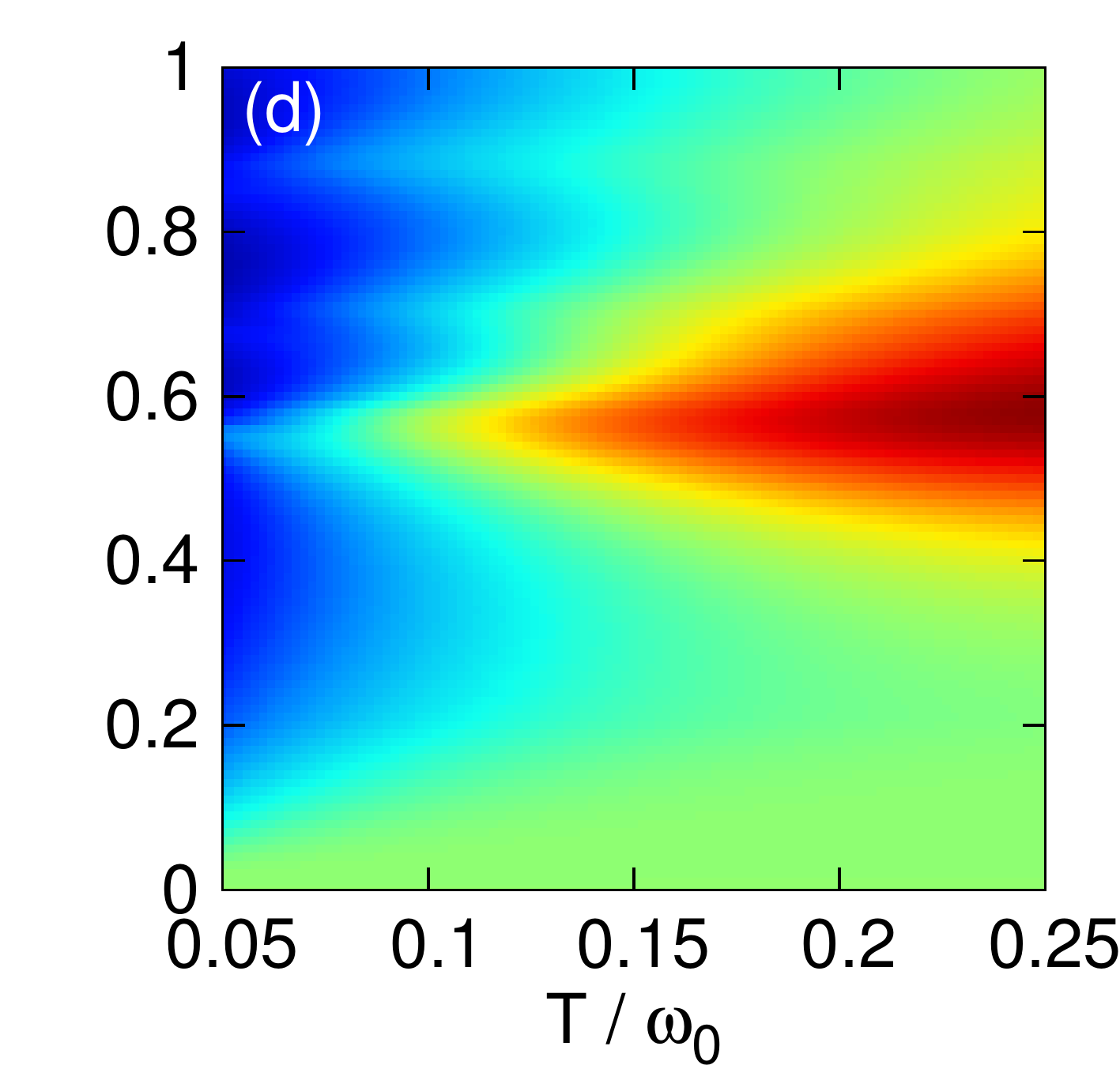}
 \caption{\label{fig:g2_chart}(Color online) Glauber function $g^{(2)}(0)$ at zero time-delay
 as a function of temperature $T$ and coupling strength $g$,
 for $N=2$ emitters (top panels a,c) and $N=3$ emitters (bottom panels b,d),
 The left panels (a,b) show results for $g'=g$, whereas $g'=0$ in panels (c,d).
  Note that all values $g^{(2)}(0) \ge 4$ are assigned the same (dark-red) color in the density plots.
 }
\end{figure}

\subsection{\label{sec:stats_scale}Photon statistics for few emitters}

The distinctive features of the Glauber function persist for multiple emitters (see Fig.~\ref{fig:g2_chart}), but the regions are shifted to smaller couplings $g$ as the number of emitters increases from one to three.

The obvious similarity between $g^{(2)}(0)$ for $N=1,2,3$ emitters visible
in Figs.~\ref{fig:g2_chart_ext},~\ref{fig:g2_chart} can be expressed
as an approximate relation between the respective emitter-cavity coupling $g$.
In the Dicke limit ($g'=g$) we find that the features of $g^{(2)}(0)$ are closely reproduced under the scaling $g \propto 1/N$.
In the TC limit ($g'=0$) features are reproduced under the scaling $g \propto 1/\sqrt N$.
Interestingly, the proper scaling of $g$ depends on the presence of counter-rotating interaction terms in the Hamiltonian.
This difference is in contrast to the semiclassical theory where the mean cavity photon number in the steady state scales $\propto N$  both in the Dicke and TC limit.
Not surprisingly, the Glauber function $g^{(2)}(0)$ is more sensitive to the details of light-matter coupling than the semiclassical theory that neglects quantum correlations in favor of a mean-field approximation.

Our arguments in favor of the above scaling relations depend on several observations, which we now develop for the TC limit ($g'=0$).
Without counter-rotating interaction terms the Hamiltonian $H$ commutes with the operator $N_t = a^\dagger a + \sum_{j=1}^N \sigma_+^{(j)} \sigma_-^{(j)}$,
which counts the total number of excitations.
Hence, $H$ is block diagonal with blocks of the form $n_t \omega_0 \mathbb{I} + g \mathbf{C}$, where $n_t$ denotes the eigenvalue of $N_t$, $\mathbb{I}$ is the identity matrix, 
and the matrix block $\mathbf C$ contains the $g$-independent matrix elements of the co-rotating interaction terms in $H$.
From this form of the blocks it is evident that the eigenvectors of $H$ do not depend on $g$,
i.e., 
the matrix elements of $\dot X_\pm$ that enter Eq.~\eqref{glauber} are constant.
The dependence of $g^{(2)}(0)$ on $g$ results from the eigenvalues only,
which determine the occupation of the states in the stationary (thermal) state and the prefactors of $\dot X_\pm$.
If we can show that the eigenvalues scale approximately as $g \sqrt N$ the above relation follows.

\begin{figure}
 \includegraphics{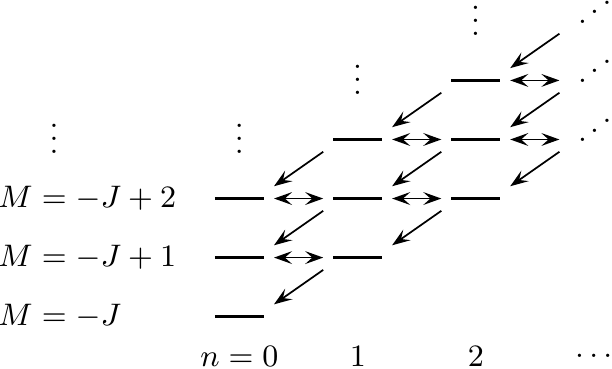}
 \caption{\label{fig:ladder_RWA_a}Schematic energy level pattern.
  Horizontal arrows depict the co-rotating interaction terms in Eq.~\eqref{HS} (coupling constant $g$), diagonal arrows illustrate the action of the cavity photon annihilation operator $a$.}
\end{figure}

Let us focus on the low lying states that give the dominant contribution in the interesting temperature regimes.
These states can be found in the ladder diagram of states in Fig.~\ref{fig:ladder_RWA_a}.
They must be connected to the ground state at energy zero by a diagonal arrow that gives the action of the operator $a$, i.e., of $\dot X_-$.

For the denominator $\langle \dot{X}_+ \dot{X}_- \rangle$ of $g^{(2)}(0)$ from Eq.~\eqref{glauber}
states contribute which are separated by one vertical step in the ladder diagram.
The energy of the most relevant first excited state is given by $E_1 = \omega_0 \pm g \sqrt N$, which has the postulated scaling.
This scaling of the first excited state for few emitters has been verified experimentally in Ref.~\cite{FBBGSFLBW09}.

For the numerator $\langle \dot{X}_+ \dot{X}_+ \dot{X}_- \dot{X}_- \rangle$ of $g^{(2)}(0)$, where each operator appears twice, states contribute which are separated by two vertical steps on the ladder.
Now the second excited state is most relevant,
which is the linear combination of the two ($N=1$) or three ($N \ge 2$) vertical rungs that occur for $n_t=2$ excitations. 
The corresponding $2\times 2$ or $3 \times 3$ matrix from the above block decomposition of $H$ is
\begin{equation}\label{mat_numer}
 \begin{pmatrix} 2 \omega_0 & \sqrt{2} g \\ \sqrt{2} g & 2 \omega_0 \end{pmatrix} \,, \quad
 \begin{pmatrix} 2 \omega_0 & \sqrt{2} g & 0 \\ \sqrt{2} g & 2 \omega_0 & \sqrt{2 N} g \\ 0 & \sqrt{2 N} g & 2 \omega_0 \end{pmatrix} \;.
\end{equation}
Diagonalization gives the energies $E_2 = 2 \omega_0 \pm \sqrt{2} g$ for $N = 1$, while $E_2 \in \{ 2 \omega_0, 2 \omega_0 \pm \sqrt{2} \sqrt{N + 1} g \}$ for $N \geq 2$.
With the approximation $\sqrt{N+1} \approx \sqrt N$,
which is good enough for a rule of thumb,
this is again the postulated scaling.
Put together, the energies that enter the computation of $g^{(2)}(0)$ 
scale roughly as $g \sqrt N$, which concludes our argument in favor of the observed relation ``$g \propto 1/\sqrt N$'' in the TC limit.

In the Dicke limit $g'=g$ the block decomposition of $H$ is not possible because of the counter-rotating interaction terms.
The eigenvectors of $H$ now depend on $g$, and the previous argument cannot be easily translated.
However, inspection of the energy spectra in Fig.~\ref{fig:E} strongly suggests that the observed relation is still related to an approximate relation between the eigenvalues of $H$ for different $N$, now with the scaling $g \propto 1/N$.

\subsection{Photon statistics from the quantum optical master equation}

\begin{figure}
 \includegraphics[width=0.49\linewidth]{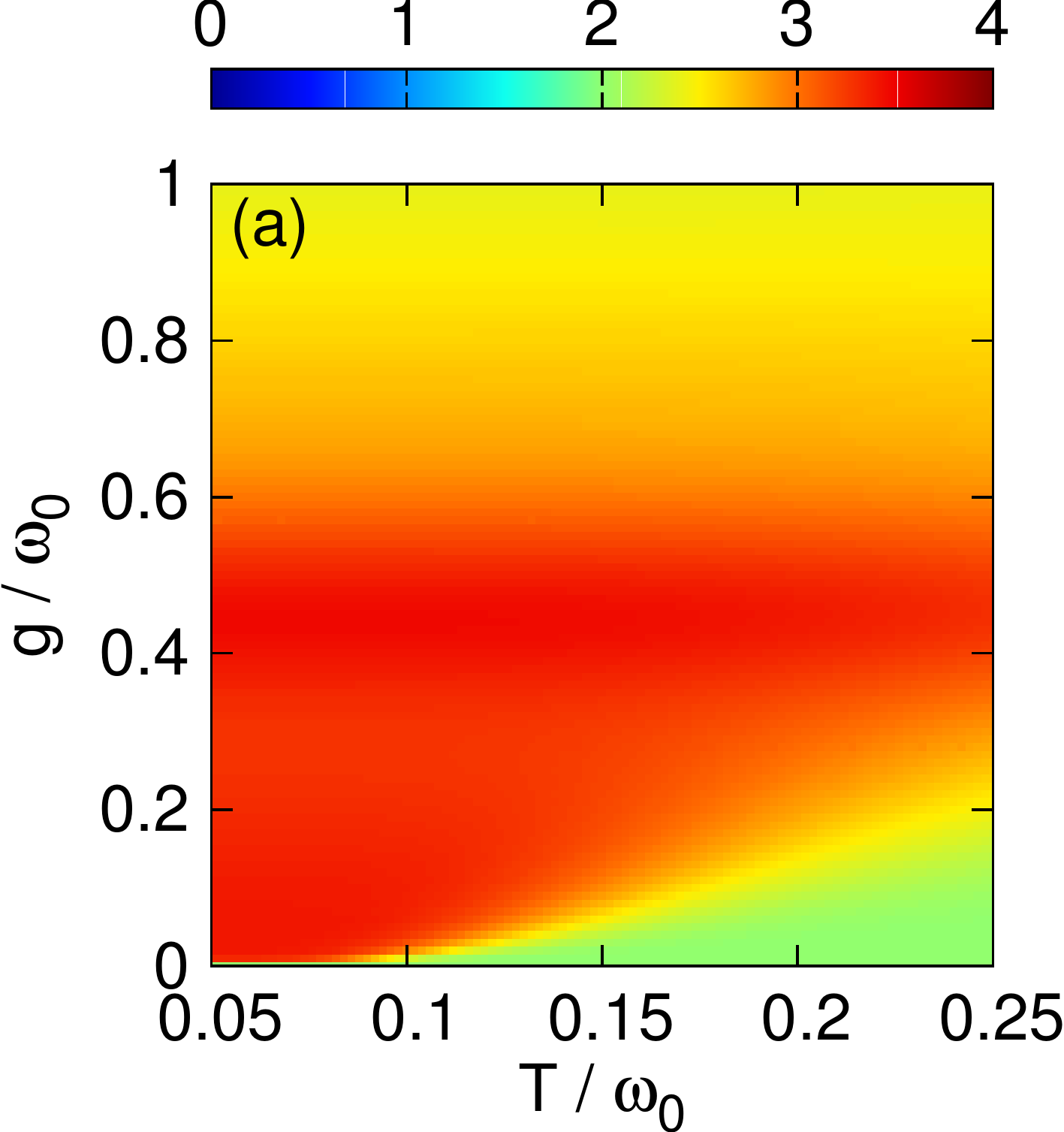}
 \includegraphics[width=0.49\linewidth]{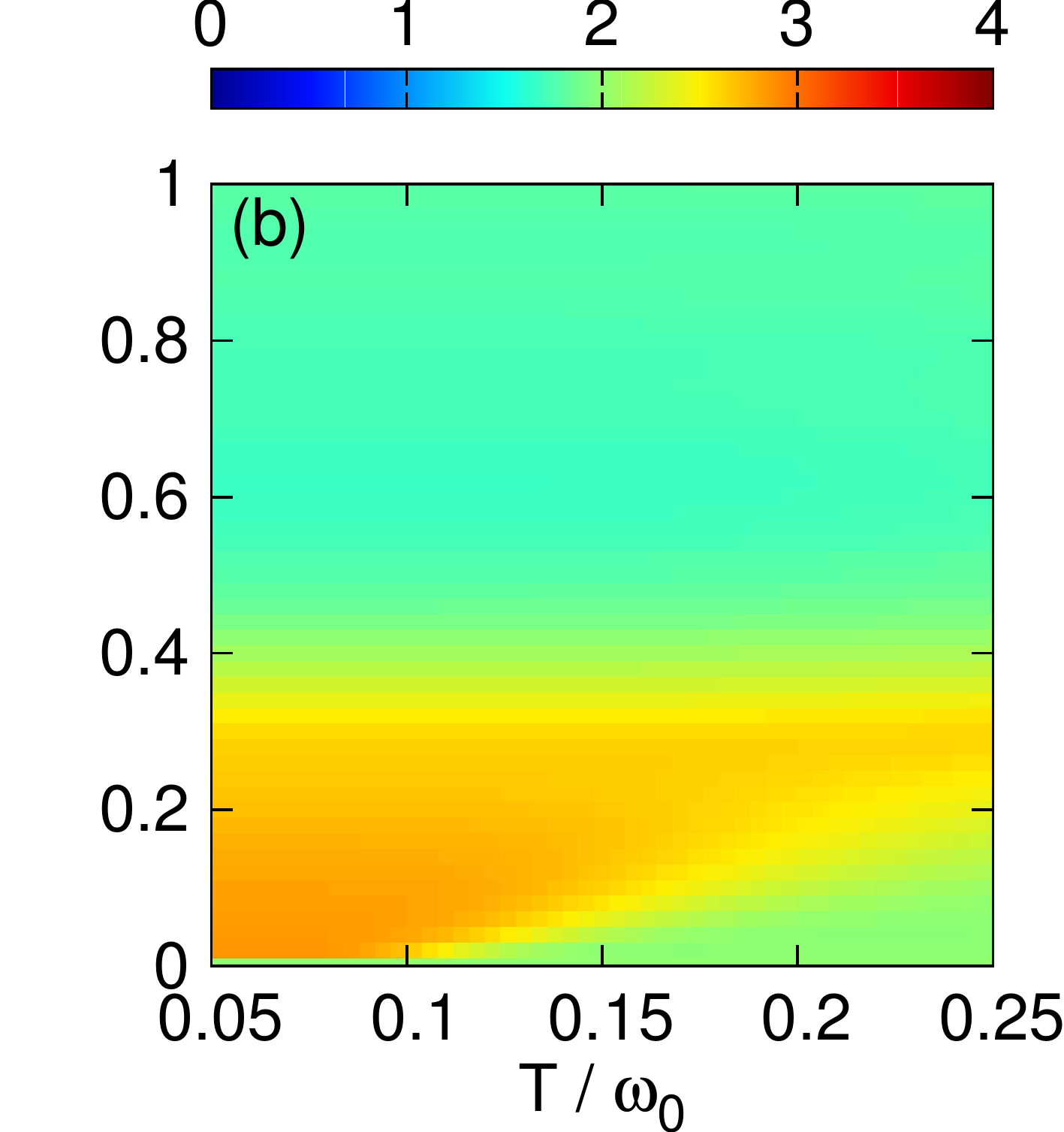}
 \caption{\label{fig:g2_chart_qo}(Color online) Glauber function $g^{(2)}(0)$
computed with the quantum optical master equation~\eqref{ME_QO}, shown as a function of temperature $T$ and coupling strength $g$ for $g'=g$.
Panel (a) gives the result for $N=1$ emitter, whereas $N=2$ in panel (b).
 }
\end{figure}

Results for the Glauber function obtained with the quantum optical master equation~\eqref{ME_QO} are shown in Fig.~\ref{fig:g2_chart_qo} in the Dicke limit $g'=g$.
In stark contrast to the results from Figs.~\ref{fig:g2_chart_ext},~\ref{fig:g2_chart} 
the quantum optical master equation does not predict the emission of nonclassical light with sub-Poissonian photon statistics  in any part of the parameter space.
The situation does not improve in the TC limit $g'=0$ where $[H, N_t] = 0$ and the quantum optical master equation gives the stationary (thermal) state $\propto \rme^{-\beta \omega_0 N_t}$ leading to $g^{(2)}(0) = \langle a^\dagger a^\dagger a a \rangle / \langle a^\dagger a \rangle^2 = 2$ independent of the number of emitters $N$, the coupling strength $g$, or the temperature $T$, thus always predicting the emission of thermal light.
While it may not be surprising that the quantum optical master equation fails, because the weak coupling condition $g \ll \omega_{x,c}$ is not satisfied, it is remarkable that it fails to capture any features from the previous Glauber function plots
in Figs.~\ref{fig:g2_chart_ext},~\ref{fig:g2_chart}.
This failure highlights the importance of using the correct master equation not only for strong light-matter coupling
but also if one is interested in properties following from higher-order correlation functions,
such as the photon statistics obtained from the second order Glauber function.

\subsection{Photon bunching and antibunching}
A further property to distinguish classical and nonclassical light is the time-coincidence statistics of the emitted photons,
which can be deduced from the time-dependent Glauber function $g^{(2)}(t)$.
For classical light, $g^{(2)}(t)$ has a non-positive initial slope at $t=0$.
This indicates photon bunching, i.e., that the probability of observing two photons at equal times is larger than the probability of observing them at different times.
Conversely, a positive slope indicates photon antibunching, which is possible only for nonclassical light.
In the long-time limit, $\lim_{t \to \infty} g^{(2)}(t) = 1$ in all cases.

\begin{figure}
 \includegraphics[width=0.49\linewidth]{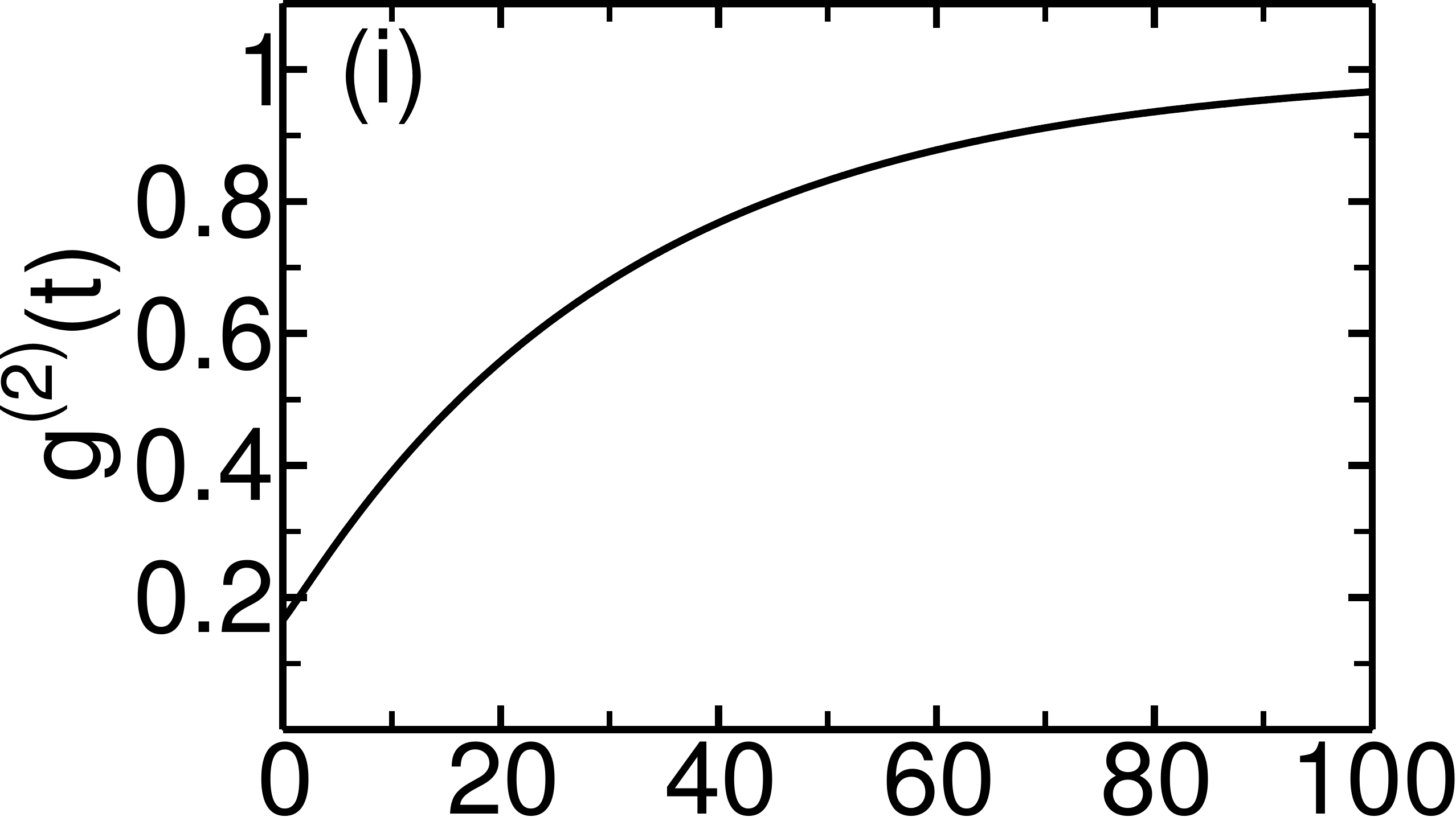}
 \includegraphics[width=0.49\linewidth]{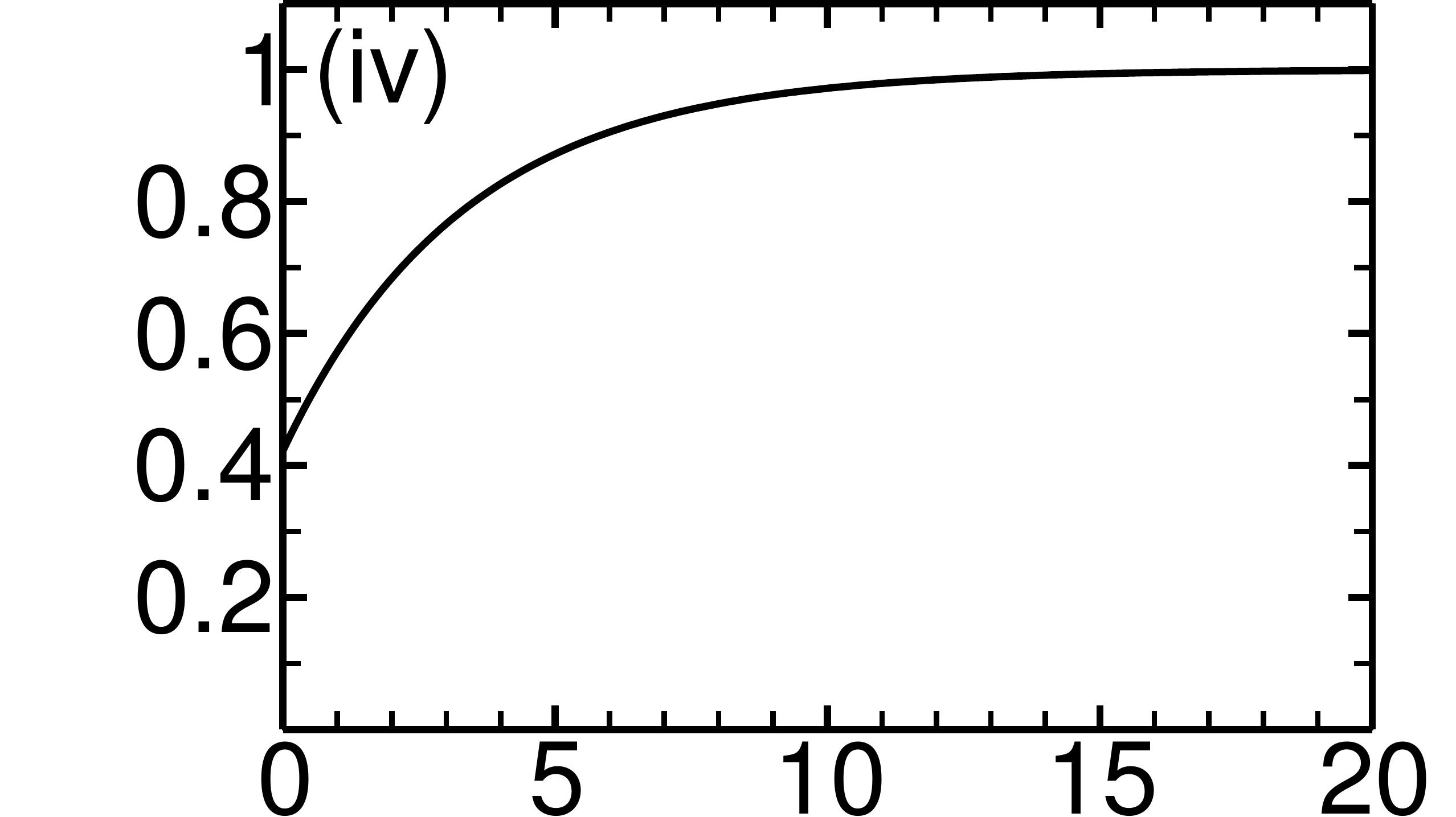}\\
 \includegraphics[width=0.49\linewidth]{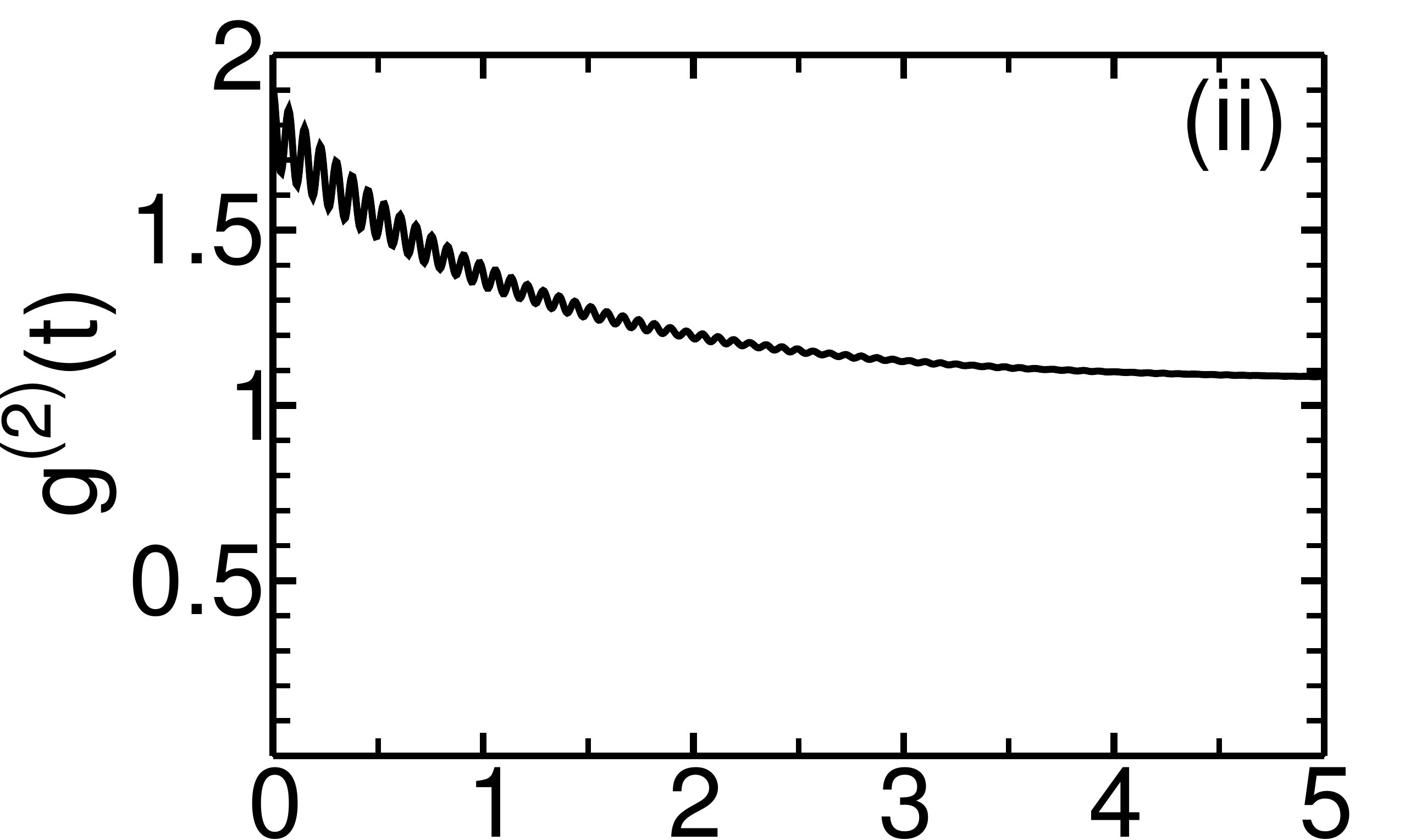}
 \includegraphics[width=0.49\linewidth]{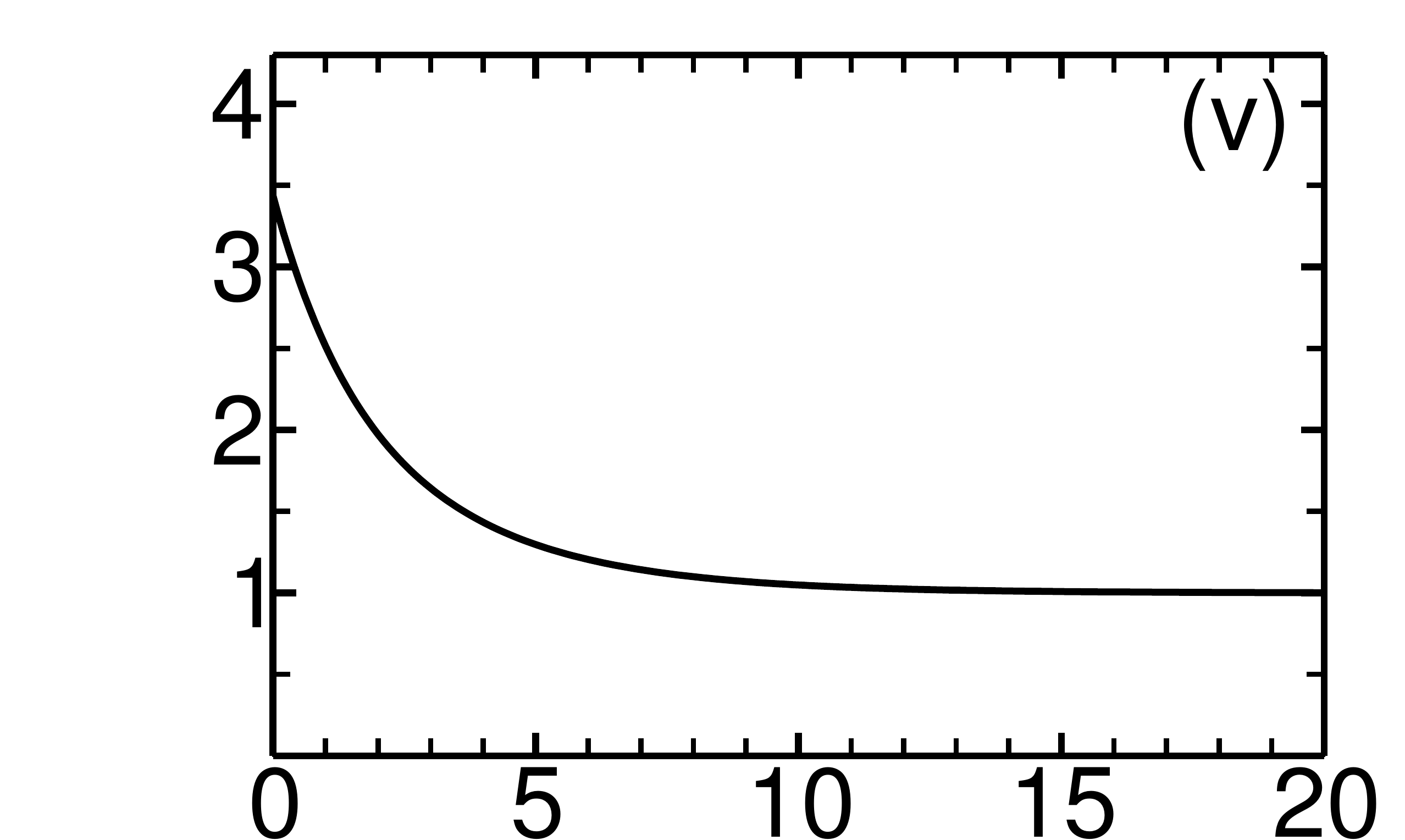}\\
 \includegraphics[width=0.49\linewidth]{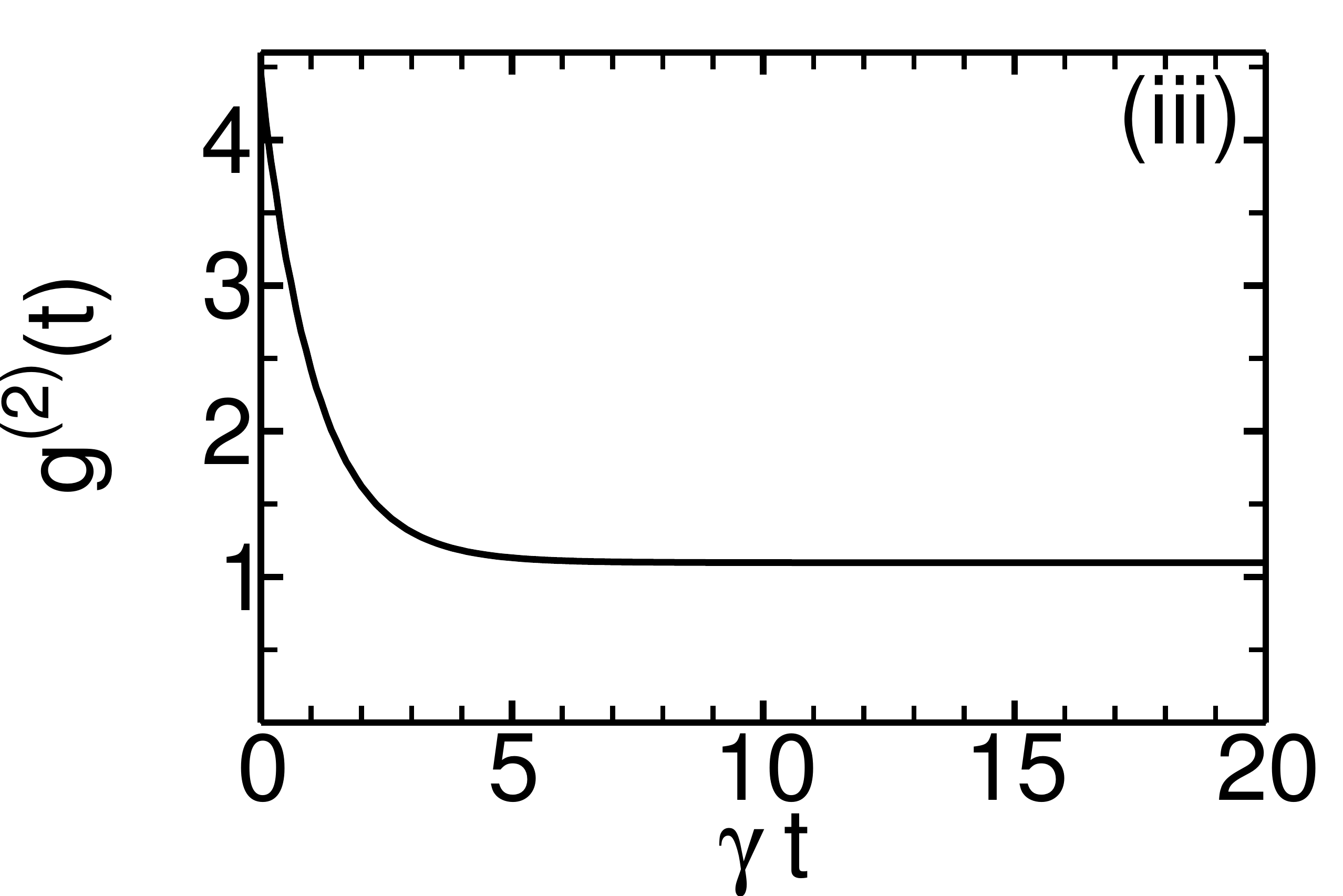}
 \includegraphics[width=0.49\linewidth]{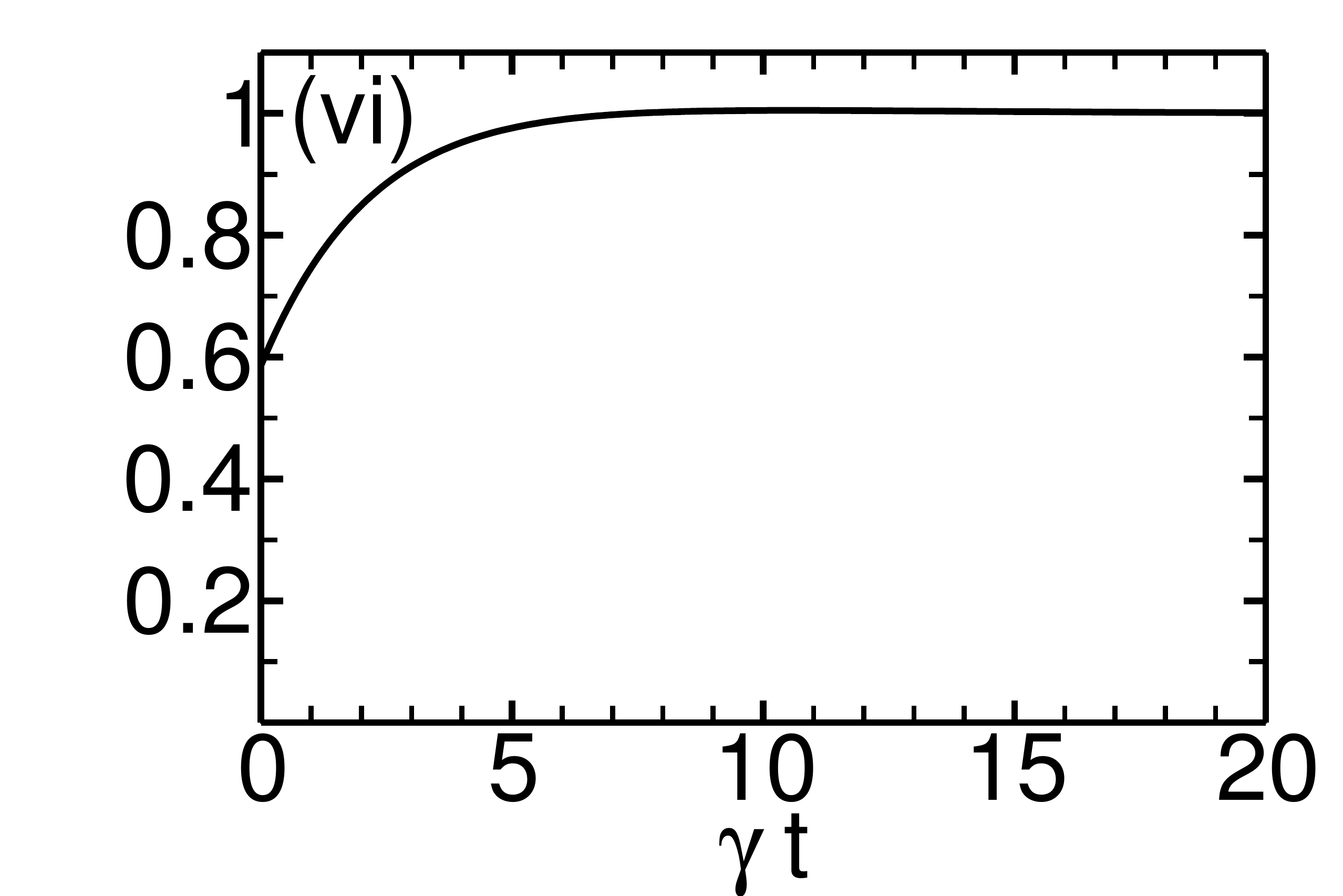}
 \caption{\label{fig:g2_time}Glauber function $g^{(2)}(t)$ as a function of time for $N=2$ emitters.
  Panels (i-iii) show the results for $g'=g$, whereas in panels (iv-vi) $g'=0$.
  The emitter-cavity coupling strength and the bath temperature are $g = 0.5 \omega_0$, $T = 0.07 \omega_0$ in panels (i, iv), $g = 0.7 \omega_0$, $T = 0.23 \omega_0$ in panels (ii,v), and $g = 0.8 \omega_0$, $T = 0.1 \omega_0$ in panels (iii,vi).}
\end{figure}

In Fig.\ \ref{fig:g2_time} we plot $g^{(2)}(t)$ for the parameter combinations marked in the two upper panels in Fig.\ \ref{fig:g2_chart}.
We see that $g^{(2)}(t)$ is always a strictly monotonic function of $t$.
Therefore, in the present situation photon bunching and antibunching coincide precisely with super-Poissonian and sub-Poissonian photon statistics.
Only if $1 \leq g^{(2)}(0) \leq 2$ in panel (ii) the function $g^{(2)}(t)$ oscillates slightly, but the overall decay is still indicative of photon bunching.

\section{\label{sec:concl}Conclusions}

Our analysis of the light generated by few emitters in a cavity
reveals a non-trivial dependence of the photon statistics
on the light-matter coupling and temperature.
Clearly identifiable parameters regimes with sub- and super-Poissonian photon statistics appear at strong and ultrastrong coupling,
and lie immediately next to each other.
Tuning of the light-matter coupling or change of the temperature can thus have a tremendous effect on the photon statistics.
As a general trend we find strong signatures of nonclassical light at strong coupling.
Thermal photon statistics, on the other hand, requires weak coupling or high temperatures: It is the exception rather than the rule at low temperatures.

The photon statistics, and to a lesser degree also the total emission,
is strongly influenced by the presence of counter-rotating light-matter interaction terms in the Hamiltonian. These terms are responsible for the prevalence of super-Poissonian over sub-Poissonian light at ultrastrong coupling.
Not surprisingly, the convenient rotating-wave approximation (i.e., identification of the Dicke by the TC limit) gives the wrong prediction when the coupling becomes too large.
Nevertheless, the scenarios with and without counter-rotating terms are surprisingly similar at not too strong coupling,
which shows that generation of nonclassical light is not a peculiar effect arising from the fine-tuning of interaction terms  in the Hamiltonian but a rather robust feature.

We have provided an approximate rule to relate the emission of few emitters to the emission of a single emitter, under appropriate scaling of the coupling constant.
In accordance with this rule, the features of the Glauber function observed for one emitter occur at comparably smaller values of the individual emitter-cavity coupling in the case of a few emitters. The reason is that all emitters interact with the same cavity mode, which magnifies the effects of resonant emission and (re-)absorption of cavity photons.
Broadly speaking, generation of nonclassical light is easier with more emitters because the required coupling of each individual emitter to the cavity mode can be reduced.

Our analysis of strong light-matter coupling required use of the full input-output formalism and of the full master equation, which carefully distinguishes between transitions at different energies.
If this correct treatment is replaced by the standard quantum optical master equation results change completely.
Especially, the prediction of nonclassical light does not survive the additional approximations made in the replacement.
While the quantum optical master equation could not be expected to work at strong coupling, its outright failure at describing any 
of the distinctive features observed in the photon statistics
shows that use of the right master equation is essential in all situations, perhaps apart from extremely weak coupling.
The price one has to pay is full diagonalization of the Hamiltonian.

We here focus on the system at thermal equilibrium.
Future work should address emission if the system is driven coherently through external photon sources.
This will require addition of explicitly time-dependent periodic terms to the Hamiltonian, and thus combination of the present master equation with the Floquet formalism.
By contrast, a perturbative expansion in the driving strength is sufficient only for weak off-resonant driving, but then the possible new effects would be weak too.

\begin{acknowledgments}

This work was supported by Deutsche Forschungsgemeinschaft through Sonderforschungsbereich 652 project B5.

\end{acknowledgments}

\appendix
\section{\label{app:inout}The input-output formalism}
We follow standard input-output theory~\cite{GZ04, RLSH12}.
The interaction Hamiltonian in Eq.\ \eqref{HI} for the cavity-environment coupling in the continuum limit is
\begin{equation}
 H_I = -\rmi X \int D(\omega) \lambda(\omega) (b_\omega - b_\omega^\dagger) \rmd\omega \,,
\end{equation}
where $D(\omega)$ is the environment density of states and $\lambda_c(\omega)$ the cavity-environment coupling function, i.e., the environment spectral function is $\gamma_c(\omega) = 2 \pi D(\omega) \lambda_c(\omega)^2$.
$H_I$ together with the free Hamiltonian $\int D(\omega) \omega b_\omega^\dagger b_\omega \rmd\omega$ of the environment photons and the commutator $[b_\omega, b_{\omega'}^\dagger] = \delta(\omega - \omega')$ lead to the equation of motion
\begin{equation}\label{b-dot}
 \dot{b}_\omega = -\rmi \omega b_\omega + \lambda(\omega) X
\end{equation}
for the field quadratures of the environment.
For $t_0 < t < t_1$, the formal solution of Eq.\ \eqref{b-dot} is
\begin{eqnarray}
 b_\omega(t) &=& \rme^{-\rmi \omega (t - t_0)} b_\omega(t_0) + \lambda(\omega) \int_{t_0}^t \rme^{-\rmi \omega (t - t')} X(t') \rmd t' \nonumber\\
 &=& \rme^{-\rmi \omega (t - t_1)} b_\omega(t_1) - \lambda(\omega) \int_t^{t_1} \rme^{-\rmi \omega (t - t')} X(t') \rmd t' \,. \nonumber\\
\end{eqnarray}
We define input (output) field operators
\begin{equation}
 b_{\text{in(out)}}(t) = \int D(\omega) \lambda(\omega) \rme^{-\rmi \omega (t - t_{0(1)})} b_\omega(t_{0(1)}) \rmd\omega
\end{equation}
and make use of the spectral function $\gamma_c(\omega) = \gamma \omega / \omega_0$ to obtain the input-output relation
\begin{equation}
 b_{\text{out}}(t) = b_{\text{in}}(t) + \rmi \frac{\gamma}{\omega_0} \dot{X}_-(t) \,,
\end{equation}
where $\dot{X}_-$ denotes the positive frequency component of $\dot{X}$, i.e., $\dot{X}_-$ acts as a lowering operator.
The explicit definition of $\dot{X}_-$ in the system-energy eigenbasis is given in Eq.\ \eqref{X}.

\section{\label{app:ME}The Markovian master equation}
We consider the dissipative dynamics of the system density matrix in the weak system-environment coupling limit.
For strong coupling within the system the quantum optical master equation predicts unphysical emission from the ground state~\cite{WDDVB08}.
Going one step back in the derivation of the quantum optical master equation, the second-order time-convolutionless projection operator method~\cite{BP02} gives a time-local master equation leading to consistent results including the counter-rotating terms~\cite{DLGCC09, CDLGC12}.
Nevertheless, this master equation does in general not generate positive dynamics~\cite{DS79, Ali89}.
This problem was resolved by a recently derived master equation in the system eigenbasis~\cite{BP02,ALZ06,SB08,BGB11,Sch14,Agar13} and we here recapitulate its derivation.

The total Hamiltonian is the sum of the contribution of the system, $H$, the contribution of the reservoir, $H_R$, and the interaction $H_I$.
We note that the interaction Hamiltonian in Eq.\ \eqref{HI} is of the general form $H_I = S R$, where $S$ ($R$) is a Hermitian system (reservoir) operator.
A more general coupling $H_I = \sum_n S_n R_n$ can also be considered, but leads to the same qualitative results.
The dynamics of the density operator $\hat{\rho}_T(t)$ of the total system in the interaction picture is described by the von Neumann equation
\begin{equation}\label{vonNeumann}
 \frac{\rmd}{\rmd t} \hat{\rho}_T(t) = -\rmi [\hat{H}_I(t), \hat{\rho}_T(t)] \,.
\end{equation}
As a notational convenience, we mark operators in the interaction picture with a hat.
The interaction Hamiltonian and the density operator in the interaction picture are defined as
\begin{eqnarray}
 \hat{\rho}_T(t) &=& U_0^\dagger(t, 0) \rho_T(t) U_0(t, 0) \,, \\
 \hat{H}_I(t) &=& U_0^\dagger(t, 0) H_I U_0(t, 0) \,,
\end{eqnarray}
where the time evolution operator of the uncoupled system and reservoir is
\begin{equation}
 U_0(t, s) = \rme^{-\rmi (H + H_R) (t - s)} \,.
\end{equation}

In the limit of weak system-reservoir coupling several approximations are performed.
First of all, within the Born approximation initial factorization of the density operator is assumed, $\rho_T(0) = \rho(0) \rho_R$, and the back-action of the system onto the reservoir is neglected, $\rho_T(t) = \rho(t) \rho_R$.
Secondly, the Markov approximation is performed by replacing $\rho(\tau)$ at retarded times $\tau$ with $\rho(t)$ at the local time $t$.
In the third place, assuming that the reservoir correlation time is small compared to the relaxation time of the system, the time integration is extended to infinity to arrive at the Born-Markov equation of motion
\begin{equation}\label{ME_BM_TrR}
 \frac{\rmd}{\rmd t} \hat{\rho}(t) = -\int_0^\infty \mathrm{Tr}_R \big\{ \big[ \hat{H}_I(t), [\hat{H}_I(t - \tau), \hat{\rho}(t) \rho_R] \big] \big\} \rmd\tau \,,
\end{equation}
where $\mathrm{Tr}_R\{\cdot\}$ denotes the partial trace over the reservoir degrees of freedom and $\langle R \rangle = 0$ is assumed.
We further assume a thermal reservoir state $\rho_R \propto \rme^{-\beta H_R}$ and define the reservoir correlation function
\begin{equation}
 C(\tau) = \mathrm{Tr}_R \big\{ \rme^{\rmi H_R \tau} R \rme^{-\rmi H_R \tau} R \rho_R \big\} = C(-\tau)^*
\end{equation}
to evaluate the traces in Eq.~\eqref{ME_BM_TrR}.
This yields the master equation
\begin{equation}
 \frac{\rmd}{\rmd t} \hat{\rho}(t) = \int_0^\infty \big[ \hat{S}(t - \tau) \hat{\rho}(t), \hat{S}(t) \big] C(\tau) \rmd\tau + \text{H.c.} \,,
\end{equation}
where H.c.\ denotes the Hermitian conjugate.

We introduce the transition operators in Eq.\ \eqref{S} that are the discrete Fourier components of the interaction picture $\hat{S}(t)$, i.e.,
\begin{equation}
 \hat{S}(t) = \sum_\omega \rme^{-\rmi \omega t} S_\omega \,.
\end{equation}
Equivalently, $[H, S_\omega] = -\omega S_\omega$.
In addition, we introduce the even and odd Fourier transforms of the reservoir correlation function
\begin{eqnarray}
 \chi(\omega) &=& \int_{-\infty}^\infty C(\tau) \rme^{\rmi \omega \tau} \rmd\tau = \chi(\omega)^* \,, \\
 \xi(\omega) &=& \frac{1}{\rmi} \int_{-\infty}^\infty C(\tau) \mathop{\mathrm{sgn}}(\tau) \rme^{\rmi \omega \tau} \rmd\tau = \xi(\omega)^* \,.
\end{eqnarray}
For a thermal photon reservoir with spectral function $\gamma(\omega)$ the functions $\chi(\omega)$ and $\xi(\omega)$ are given in Eqs.\ \eqref{chi} and\ \eqref{xi}.
With these definitions we find
\begin{eqnarray}\label{ME_BM}
 \frac{\rmd}{\rmd t} \hat{\rho}(t) &=& \frac{1}{2} \sum_{\omega, \omega'} \big\{ \chi(\omega') + \rmi \xi(\omega') \big\} \rme^{\rmi (\omega - \omega') t} \big[ S_{\omega'} \hat{\rho}(t), S_\omega^\dagger \big] \nonumber\\
 && + \text{H.c.} \,.
\end{eqnarray}
Eq.\ \eqref{ME_BM} is the standard Born-Markov master equation in the system energy-eigenbasis.
It contains the dissipative parts proportional to $\chi(\omega)$ and the Lamb-shift terms proportional to $\xi(\omega)$.
Because Eq.\ \eqref{ME_BM} is not of Lindblad type, it does, in general, not preserve the positivity of the density operator.

Inspecting Eq.\ \eqref{ME_BM} we recognize that it contains oscillating terms proportional to $\rme^{\pm \rmi (\omega - \omega') t}$.
If we assume that the relaxation of the system is slow compared with all oscillations $\rme^{\pm \rmi (\omega - \omega') t}$ we can neglect the contribution from terms with $\omega' \neq \omega$.
This approximation is called secular or rotating-wave approximation and the master equation in the Schr\"{o}dinger picture simplifies to the result given in Eq.\ \eqref{ME}.
This equation is the Lindblad master equation that includes the Lamb shift of the unperturbed system energies $E_n$ as well as reservoir induced dissipation effects to lowest order in the system-reservoir interaction strength.

As is already known in the literature, special care has to be taken if the spectrum of $H$ is degenerate \cite{JS04, BGB11}.
But even if the eigenvalues $E_n$ are non-degenerate we may have situations where energy differences are degenerate, i.e.\ $E_n - E_m = E_k - E_l$ for $n \neq m \neq k \neq l$.
The consequences of these two different types of degeneracy can be understood when we decompose the density matrix into blocks.
In particular, we write $\hat{\boldsymbol{\rho}}_{nm}$ ($\mathbf{S}_{mn}$) for the matrix containing the elements $\langle k | \hat{\rho} | l \rangle$ ($\langle k | S | l \rangle$) with $E_k = E_n$ and $E_l = E_m$.
The master equation\ \eqref{ME} in this block notation reads
\begin{eqnarray}
 \frac{\rmd}{\rmd t} \hat{\boldsymbol{\rho}}_{mn}(t) &=& \sum_{k,l} \chi(E_k - E_m) \mathbf{S}_{mk} \hat{\boldsymbol{\rho}}_{kl}(t) \mathbf{S}_{ln}^\dagger \delta_{E_k - E_m, E_l - E_n} \nonumber\\
 && - \frac{1}{2} \sum_k \varphi(E_n - E_k)^* \hat{\boldsymbol{\rho}}_{mn}(t) \mathbf{S}_{nk}^\dagger \mathbf{S}_{kn} \nonumber\\
 && - \frac{1}{2} \sum_k \varphi(E_m - E_k) \mathbf{S}_{mk}^\dagger \mathbf{S}_{km} \hat{\boldsymbol{\rho}}_{mn}(t) \,,
\end{eqnarray}
where the summations are over different system energies only, and the complex function $\varphi(x) = \chi(x) + \rmi \xi(x)$ is introduced.
We see that the last two lines in this equation are block-diagonal.
For $m = n$, the Kronecker-delta in the first line evaluates to $\delta_{E_k, E_l}$ such that diagonal blocks couple to diagonal blocks, only.
For $m \neq n$, the first line contains terms with $k \neq l$ only, such that non-diagonal blocks do not couple to diagonal ones.
Nevertheless, a non-diagonal block $\hat{\boldsymbol{\rho}}_{mn}$ couples to another non-diagonal block $\hat{\boldsymbol{\rho}}_{kl}$ with $k \neq l \neq m \neq n$ if the respective transition energies are degenerate.
Thus, energy level degeneracy introduces a block structure implying that a diagonal density matrix element couples to non-diagonal elements within diagonal blocks whereas energy transition degeneracy leads to a coupling of non-diagonal blocks to different non-diagonal blocks.
We remark that both subtleties have their origin in the rotating wave approximation.
On the one hand, this approximation leads to the Lindblad structure of Eq.\ \eqref{ME}.
On the other hand, it results in strict Kronecker delta's between the two transition energies $\omega'$ and $\omega$.

Consider a situation where each degeneracy in the spectrum of $H$ as well as in their differences is lifted by a small $\epsilon$ parameter.
Then, each block contains a single element only, implying that the equations for the diagonal density matrix elements no longer couple to non-diagonal elements.
In addition, any non-diagonal element of the density matrix evolves independently from all other elements.
This behavior does not change when we let each $\epsilon \to 0$.
In this limit, the equations become independent of the $\epsilon$ parameters, but are different from the $\epsilon = 0$ case.
In particular, for every non-zero $\epsilon \to 0$ we get the two equations\ \eqref{ME_nn} and\ \eqref{ME_mn} for the diagonal and non-diagonal density matrix elements.

We remark that in real physical systems one will never have perfectly equal or equidistant energies because each small perturbation will lift the degeneracies.
In the theoretical description we may argue that the Lamb shift lifts degeneracies.
Nevertheless, we have to keep in mind, that with Eqs.\ \eqref{ME_nn} to\ \eqref{Zn} we can not study effects that rely on degenerate energies or degenerate transitions, e.g.\ the perfectly harmonic oscillator or a system composed of completely uncoupled identical subsystems are not correctly described.

\section{\label{app:analyt}Analytical results in the Tavis-Cummings limit}
In this section we derive analytical results for the Glauber $g^{(2)}(0)$-function in the TC limit ($g' = 0$) for a single emitter ($N = 1$).

According to the argumentation in Sec.\ \ref{sec:stats_scale}, the dominant contribution to the denominator of $g^{(2)}(0)$ in Eq.\ \eqref{glauber} at low temperatures is that of the first excited state with energy $E_1 = \omega_0 - g$.
Specifically, the denominator $\langle \dot{X}_+ \dot{X}_- \rangle$ is approximated by
\begin{equation}\label{app:denom}
 \frac{1}{2} (\omega_0 - g)^2 \rme^{-\beta (\omega_0 - g)} \;.
\end{equation}
In this expression the exponential $\rme^{-\beta (\omega_0 - g)}$ is the thermal population of the first excited state and the prefactor $(\omega_0 - g)^2 / 2$ is the squared transition matrix element of $\dot{X}_-$ between the first excited state and the ground state.

The most relevant state for the numerator 
of $g^{(2)}(0)$ at low temperatures is the lowest eigenstate with energy $E_2 = 2 \omega_0 - \sqrt{2} g$ of the $2 \times 2$ matrix given in Eq.\ \eqref{mat_numer}.
To evaluate the matrix elements of the operators $\dot{X}_\pm$ we have to consider the four possible transition sequences $| 2, - \rangle \rightarrow | 1, \pm \rangle \rightarrow | 0 \rangle \rightarrow | 1, \pm \rangle \rightarrow | 2, - \rangle$, where $|0\rangle$ denotes the ground state and $|n,\pm\rangle$ are the two eigenstates with energies $E_n = n \omega_0 \pm \sqrt{n} g$.
This yields the expression
\begin{eqnarray}\label{app:numer}
 && \bigg\{ \frac{3 + \sqrt{8}}{8} \big[ \omega_0 - (\sqrt{2} - 1) g \big]^2 (\omega_0 - g)^2 \nonumber\\
 && + \frac{1}{4} \big[ \omega_0 - (\sqrt{2} - 1) g \big] (\omega_0^2 - g^2) \big[ \omega_0 - (\sqrt{2} + 1) g \big] \nonumber\\
 && + \frac{3 - \sqrt{8}}{8} \big[ \omega_0 - (\sqrt{2} + 1) g \big]^2 (\omega_0 + g)^2 \bigg\} \rme^{-\beta (2 \omega_0 - \sqrt{2} g)} \nonumber\\
\end{eqnarray}
approximating the numerator of $g^{(2)}(0)$.

\begin{figure}[t]
 \includegraphics[width=0.49\linewidth]{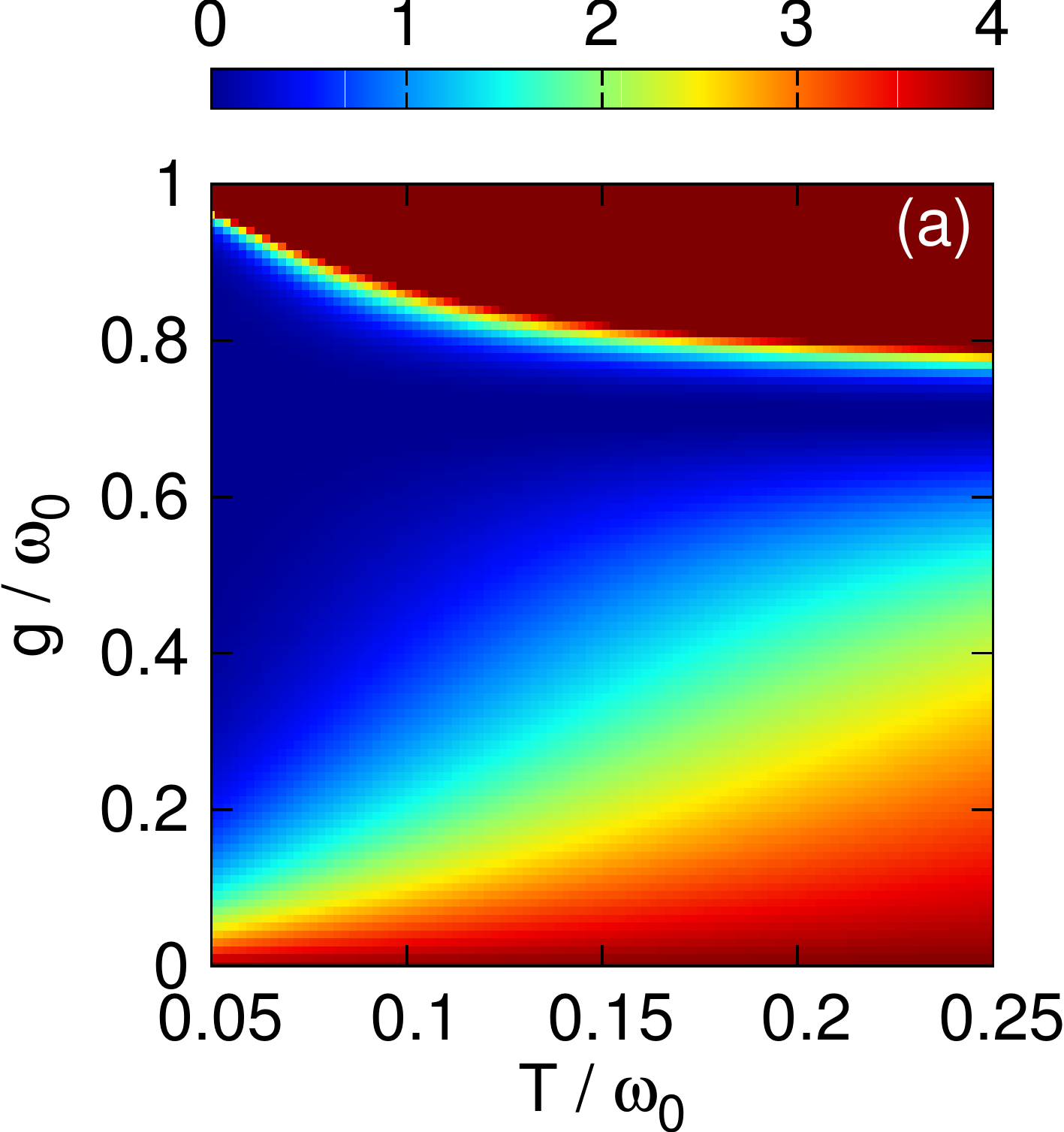}
 \includegraphics[width=0.49\linewidth]{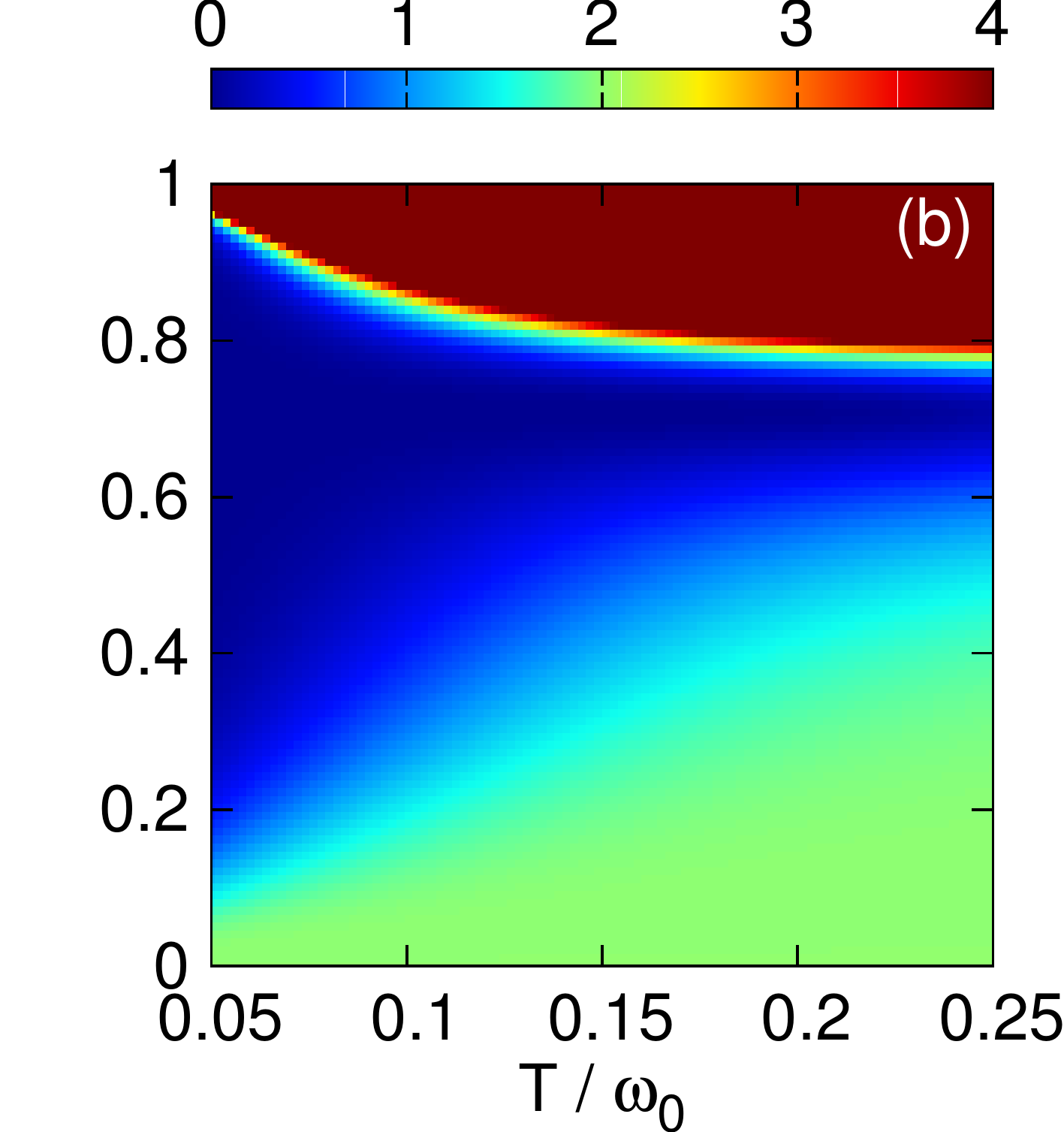}
 \caption{\label{fig:g2_chart_ana}(Color online) Analytical results for the Glauber function $g^{(2)}(0)$ for one emitter ($N=1$), as a function of temperature $T$ and coupling strength $g$ in the TC limit.
  Panel (a) shows the result following from Eqs.\ \eqref{app:denom} and\ \eqref{app:numer}, whereas in panel (b) the result is refined for $g \ll \omega_0$ as explained in the text.}
\end{figure}

The results belonging to Eqs.\ \eqref{app:denom} and\ \eqref{app:numer} are plotted in Fig.\ \ref{fig:g2_chart_ana}(a).
Compared to the exact numerical results in Fig.\ \ref{fig:g2_chart_ext}(b) a good agreement appears for $0.2 \lesssim g / \omega_0 < 0.4$ and low temperatures.
For high values of $g \gtrsim 0.4 \omega_0$ the first excited state becomes closer and closer to the ground state such that the finite temperature leads to significant contributions from transitions not involving the ground state.
For this reason, the upper part of Fig.\ \ref{fig:g2_chart_ana}(a) is not well reproduced.
In contrast, the lower part of Fig.\ \ref{fig:g2_chart_ana}(a) is not in accordance with the exact numerical results because our assumption of low temperatures $T \ll g < \omega_0$ does not include the limit $g \to 0$.

To improve the results in regions with $g \ll \omega_0$ we additionally have to take into account the transition sequences $| 1, + \rangle \rightarrow | 0 \rangle \rightarrow | 1, + \rangle$ for the denominator and $|2, + \rangle \rightarrow | 1, \pm \rangle \rightarrow | 0 \rangle \rightarrow | 1, \pm \rangle \rightarrow | 2, + \rangle$ for the numerator.
The result is shown in Fig.\ \ref{fig:g2_chart_ana}(b), where the agreement to the exact results in Fig.\ \ref{fig:g2_chart_ext}(b) is now very good for all temperatures and $g \lesssim 0.4 \omega_0$.
Note that at $g \simeq 0.4 \omega_0$ a crossing of the eigenvalues of the second and third excited state occurs, as can be seen in Fig.\ \ref{fig:E}(d).
This indicates that the role of these states in the calculation of $g^{(2)}(0)$ is interchanged.
The impact of these eigenvalue-crossings would analytically be reproduced if we include contributions to $g^{(2)}(0)$ from higher excited states.


%

\end{document}